\UseRawInputEncoding
\documentclass[1p,authoryear]{elsarticle}

\usepackage{lmodern}
\usepackage[T1]{fontenc}
\usepackage{amsmath}
\usepackage{amsthm}
\usepackage{amsfonts}
\usepackage{amssymb}
\usepackage{algorithm}
\usepackage[noend]{algpseudocode}
\usepackage{hyperref}
\usepackage{listings}
\usepackage[scaled=0.8]{beramono}
\usepackage{textcomp}

\newcommand{\comment}[1]{}
\newcommand{\RR}{\mathfrak{R}}
\newcommand{\KK}{\mathfrak{K}}
\newcommand{\TT}{\mathfrak{T}}
\newcommand{\RRm}{Rm}
\newcommand{\sig}{\ensuremath{\mathfrak{s}}}
\newcommand{\lp}{\ensuremath{\mathrm{lp}}}
\newcommand{\lc}{\ensuremath{\mathrm{lc}}}
\newcommand{\coeff}{\ensuremath{\mathrm{coeff}}}
\newcommand{\supp}{\ensuremath{\mathrm{supp}}}
\newcommand{\preceqt}{\preceq_\mathsf{t}}
\newcommand{\prect}{\prec_\mathsf{t}}
\newcommand{\dvdt}{\ensuremath{\mathsf{dvd}_\mathsf{t}}}
\newcommand{\preceqpot}{\preceq_\mathsf{pot}}

\newcommand{\poly}{poly}% could be redefined to rep-list
\newcommand{\polymapping}[1]{\ensuremath{#1\Rightarrow_0\beta}}
\newcommand{\fs}{\ensuremath{f\!s}}
\newcommand{\red}[3]{\ensuremath{\overset{#1,#2}{\longrightarrow}_#3}}

\newcommand{\rw}{\ensuremath{\trianglelefteq}}
\newcommand{\rwrat}{\ensuremath{\rw_\mathsf{rat}}}
\newcommand{\rwadd}{\ensuremath{\rw_\mathsf{add}}}
\newcommand{\spair}{\ensuremath{\mathrm{spair}}}
\newcommand{\lcm}{\ensuremath{\mathrm{lcm}}}
\newcommand{\wrt}{w.\,r.\,t.}

\newcommand{\Let}[2]{\State #1 $\gets$ #2}
\makeatletter
\newcommand{\numberedquote}[2]{%
  \protected@edef\@currentlabelname{(#1)}%
  \refstepcounter{equation}
  \indent\parbox{0.8\textwidth}{\bigskip #2}\hfill(#1)\bigskip%
}
\makeatother

\theoremstyle{plain}
\newtheorem{thm}{Theorem}[section]
\newtheorem{lem}{Lemma}[section]

\newtheorem{prop}{Proposition}[section]

\theoremstyle{definition}

\newdefinition{dfn}{Definition}[section]
\newdefinition{notation}{Notation}[section]
\newdefinition{ex}{Example}[section]
\newdefinition{rem}{Remark}[section]
\newdefinition{prob}{Problem}[section]

\journal{Elsevier}
\date{\tiny \copyright 2020. Licensed under CC-BY-NC-ND 4.0 \url{http://creativecommons.org/licenses/by-nc-nd/4.0/}}

\begin{document}
\begin{frontmatter}

\title{A Generic and Executable Formalization of\\Signature-Based Gr\"obner Basis Algorithms}

\author{Alexander Maletzky\fnref{ack}}
\ead{alexander.maletzky@risc-software.at}
\address{Research Institute for Symbolic Computation (RISC), Johannes Kepler University Linz,\\ 
Altenberger Strasse 69, A-4040 Linz, Austria\\and\\RISC Software GmbH, Softwarepark 35, A-4232 Hagenberg im M\"uhlkreis, Austria}
\fntext[ack]{The research was funded by the Austrian Science Fund (FWF): P 29498-N31.}

\begin{abstract}
We present a generic and executable formalization of signature-based algorithms (such as Faug\`ere's 
$F_5$) for computing Gr\"obner bases, as well as their mathematical background, in the Isabelle/HOL 
proof assistant. Said algorithms are currently the best known algorithms for computing Gr\"obner 
bases in terms of computational efficiency. The formal development attempts to be 
as generic as possible, generalizing most known variants of signature-based algorithms, but 
at the same time the implemented functions are effectively executable on concrete input for 
efficiently computing mechanically verified Gr\"obner bases. Besides correctness the 
formalization also proves that under certain conditions the algorithms a-priori detect and avoid all 
useless reductions to zero, and return minimal signature Gr\"obner bases.

To the best of our knowledge, the formalization presented here is the only formalization of
signature-based Gr\"obner basis algorithms in existence so far.
\end{abstract}


\end{frontmatter}

% TODO:
% - Make sure DOIs are printed.
% - Make sure there is space between journal and note in AFP references.
% - Make sure that "References" is not printed twice.

\section{Introduction}
\label{sec::Introduction}
% 2 pages

Gr\"obner bases, introduced by~\cite{Buchberger1965}, are a ubiquitous tool in computer algebra and 
beyond, as they allow to effectively solve many problems related to multivariate polynomial rings 
and ideals. Finding Gr\"obner bases is a computationally difficult task, and therefore many 
researchers have attempted to design more and more efficient algorithms over the years. This, 
finally, lead to the first \emph{signature-based algorithm}, the $F_5$ algorithm invented 
by~\cite{Faugere2002}. Nowadays, $F_5$ and its relatives are the most efficient algorithms for 
computing Gr\"obner bases, implemented in many modern computer algebra systems.

The work presented in this paper focuses on yet another implementation of signature-based 
algorithms, but this time not in a computer algebra system but in the proof assistant Isabelle/HOL 
(\cite{Nipkow2002}). The distinctive feature of our implementation is its \emph{formal
verification} by the inference kernel of Isabelle. This, of course, necessitated formalizing also the 
vast theory behind signature-based algorithms, eventually leading to an \emph{extensive}, 
\emph{generic} and \emph{executable} formalization of an important topic in 
modern computer algebra. Even more, it is the---to the best of our knowledge---first-ever 
formalization of this theory in \emph{any} proof assistant. As such, it constitutes the ultimate 
certificate that the existing informal theory is indeed correct, without even the slightest mistake 
or overlooked gap.

In the remainder we assume familiarity with the basics of Gr\"obner bases theory, including polynomial 
reduction, S-polynomials, the definition of Gr\"obner bases, and Buchberger's algorithm. Although 
we present the key definitions, theorems and algorithms of the signature-based approach to 
Gr\"obner bases in this paper, readers totally new to the subject might also want to have a glance 
at the excellent survey article by~\cite{Eder2017}, which we took as the template for our 
formalization. We must also stress that besides Faug\`ere, many more researchers have worked on 
signature-based algorithms, 
resulting in a great variety of such algorithms. Giving an exhaustive overview of \emph{all} 
variations of signature-based algorithms in existence is out of scope here, though; see 
again~\cite{Eder2017} instead.

The motivation and distinctive feature of signature-based algorithms is detecting and avoiding many 
more useless zero-reductions while computing Gr\"obner bases than other algorithms---and in some 
cases even \emph{all} useless reductions. This saves a lot of computation time and thus leads to a 
drastic speed-up. How all this relates to signatures, and what signatures are in the first place, 
will be explained in Sections~\ref{sec::Preliminaries}--\ref{sec::RewriteBases}.

The main motivation for the formalization was to verify a state-of-the-art 
algorithm for computing Gr\"obner bases. That task could be expected to be challenging due to 
the inherent complexity of the underlying mathematical theory, illustrated by the fact that 
termination of the original $F_5$ algorithm was an open problem for a decade until it was settled 
by~\cite{Galkin2012}. Our formal development of the theory may also serve as the basis for further 
theoretical investigations, e.\,g.\ by implementing, testing and verifying new improvements of the 
formalized algorithms, as well as for further formalizations in the vast area of Gr\"obner bases.

Summarizing, the key features of the work presented here are as follows:
\begin{itemize}
  \item The formalization is generic, in the sense that we consider \emph{rewrite 
bases} and allow for arbitrary term orders and rewrite orders (Section~\ref{sec::RewriteBases}). 
According to~\cite{Eder2017} this set-up covers most, if not all, existing variations of 
signature-based algorithms.

  \item All algorithms are proved to be totally correct \wrt\ their specifications. In particular, 
the algorithm for computing rewrite bases (Algorithm~\ref{alg::RB}) is shown to terminate 
for every input (Section~\ref{sec::Termination}).
  
  \item Besides correctness, we also prove that under certain conditions the algorithm indeed 
avoids \emph{all} useless zero-reductions, and that with a particular choice of the rewrite order 
it returns minimal signature Gr\"obner bases (Section~\ref{sec::Optimality}).

  \item All formally verified algorithms are efficiently executable on concrete input. `Efficient' 
means that, for instance, the algorithms operate only on \emph{sig-poly-pairs} rather 
than full module elements (Section~\ref{sec::Computations}).
\end{itemize}
The entire formalization is freely available online (\cite{Maletzky2018b}), as an entry of Isabelle's \emph{Archive of Formal Proofs} 
(AFP).\footnote{\url{http://www.isa-afp.org}}

% Note that this guarantees its logical soundness, because the AFP only accepts correct and complete (in the sense that no proofs are  missing) entries.

\subsection{Organization of the Paper}
\label{sec::Organization}

The rest of the paper is organized as follows: Section~\ref{sec::Isabelle} gives a brief overview 
of Isabelle/HOL, to make the paper as self-contained as possible. Section~\ref{sec::Preliminaries} 
presents the preliminaries of signature-based algorithms, Section~\ref{sec::Reduction} introduces 
\sig-reduction and signature Gr\"obner bases, Section~\ref{sec::RewriteBases} defines rewrite bases 
and S-pairs and establishes the connection between them, Section~\ref{sec::Algorithms} presents the 
main algorithms and hints why they are totally correct, and Section~\ref{sec::Optimality} contains 
two results concerning the `optimality' of the algorithms. Each of these sections first presents 
the various concepts and theorems in common mathematical style, before showing how they are 
formalized in Isabelle/HOL.

Section~\ref{sec::Computations}, then, explains how the formalized algorithms can be executed on 
concrete input and provides a comparison of the running times of these algorithms to other 
algorithms implemented in Isabelle/HOL and in \textit{Mathematica}. Section~\ref{sec::Conclusion}, 
finally, concludes the paper by giving quantitative information on the formalization effort 
and listing related and future work.

\section{Brief Overview of Isabelle/HOL}
\label{sec::Isabelle}

The purpose of this section is to give a brief overview of the most important aspects of 
Isabelle/HOL that are necessary for understanding the rest of the paper. Further information and 
documentation can be found in~\cite{Paulson1994,Nipkow2002,Wenzel2018} and on the Isabelle 
homepage\footnote{\url{http://isabelle.in.tum.de}}. Readers already familiar with Isabelle can safely skip 
this section.

Isabelle is a \emph{generic proof assistant}: it serves as a framework for implementing different 
object logics, such as first-order logic or higher-order logic, in one single system. As such, it 
provides the basic infrastructure needed for automated and interactive theorem proving in general: 
a small inference kernel based on higher-order unification, theory- and proof contexts, a document 
preparation interface, and many more. Isabelle/HOL is a concrete object logic implemented in 
Isabelle, namely classical higher-order predicate logic. Being the most actively developed object 
logic of Isabelle, it comes with a library of hundreds of useful mathematical concepts, such as 
numbers, sets, lists, abstract algebraic structures, etc., which new formalizations can build upon.

Formalizing a mathematical theory in Isabelle/HOL normally proceeds by \emph{definitional theory 
extensions}: new concepts are defined, properties of these concepts and their relation to existing 
concepts are proved, and so on; arbitrary axiomatizations, though possible in principle, are 
usually avoided to eliminate the risk of introducing inconsistencies to the theory. Our 
formalization goes without any such axiomatizations.

\subsection{Definitions}
\label{sec::IsaDefs}

New constants can be introduced either via explicit non-recursive definitions or as recursive functions. A simple example of the former is the following:
\begin{lstlisting}
definition subset-eq :: $\alpha$ set $\Rightarrow$ $\alpha$ set $\Rightarrow$ bool (infix $\subseteq$ 50)
  where subset-eq A B $\longleftrightarrow$ ($\forall$a$\in$A. a $\in$ B)
\end{lstlisting}
This definition introduces a new constant, \textsf{subset-eq}, of type $\alpha\ 
\textsf{set}\Rightarrow\alpha\ \textsf{set}\Rightarrow\textsf{bool}$. That means, it is a function 
taking two sets of element-type $\alpha$ as arguments and returning a boolean value. Greek letters like $\alpha$ always denote \emph{type variables}; hence, \textsf{subset-eq} is a 
\emph{polymorphic} function that cannot only be applied to sets of a particular element-type, but 
to all sets. The type constructors \textsf{set} and \textsf{bool} are built into Isabelle/HOL. The 
`infix' clause following the type is optional and instructs Isabelle to record the short infix 
notation $\subseteq$ for \textsf{subset-eq}. The actual definition of \textsf{subset-eq} comes 
after the `where' keyword: \textsf{subset-eq} holds for two arguments $A$ and $B$ if, and only if, 
every element $a$ of $A$ is also an element of $B$. Note that free variables in definitions, theorems, etc. are implicitly universally quantified, and that $\forall$ and $\in$ are built-in constants with 
the usual meaning. Therefore, \textsf{subset-eq} is indeed the usual subset 
relation; of course, it is a built-in constant of Isabelle/HOL, too.

As can be seen, Isabelle uses \emph{Curried notation} for denoting function application: 
in the definition, \textsf{subset-eq} is applied to $A$ and $B$ by mere juxtaposition, without 
parentheses. Parentheses only become necessary in nested function applications, as in $f\ (g\ x)$. 
What can also be seen in the above definition is that single arrows are used to denote equivalences 
and implications: $\longrightarrow$ for logical implication, $\longleftrightarrow$ for equivalence.

\begin{notation}
Within enclosing informal text, we will adopt the standard mathematical notation for function 
application when writing Isabelle code. For instance, we shall write $\textsf{subset-eq}(A,B)$ 
rather than $\textsf{subset-eq}\ A\ B$, because the latter does not fit very well with informal 
text. Furthermore, names of constants will be typeset in \textsf{sans serif} font to 
distinguish them from variables, which will be typeset in $italics$, as usual.
\end{notation}

As an example of a recursive definition, consider the following:
\begin{lstlisting}
function set :: $\alpha$ list $\Rightarrow$ $\alpha$ set where
  set [] = {} |
  set (x # xs) = {x} $\cup$ set xs
\end{lstlisting}
This command defines a polymorphic function \textsf{set} which maps lists (built-in type 
constructor \textsf{list}) to the set of their elements, which is achieved by structural recursion 
on the shape of the argument: if it is empty, the empty set is returned; otherwise, the list 
consists of a head $x$ and a tail $xs$, in which case \textsf{set} is applied to $xs$ recursively 
and $x$ is added to the result. Recursively defined functions must always be shown to terminate, 
to avoid potential inconsistencies. In some cases, Isabelle can do the termination proofs 
itself, whereas in more difficult situations the user has to construct the proofs interactively.

\subsection{Theorems and Proofs}
\label{sec::IsaThms}

Theorems can be stated using the synonymous `lemma', `theorem' or `corollary' keywords. For 
instance, a lemma expressing that the set of elements of the concatenation of two lists equals the 
union of the individual sets could be stated as follows:
\begin{lstlisting}
lemma set-append: set (xs @ ys) = set xs $\cup$ set ys
\end{lstlisting}
Here, \textsf{@} is infix syntax denoting the concatenation of lists $xs$ and $ys$, and recall from 
above that the two free variables $xs$ and $ys$ are implicitly universally quantified. So far, 
however, the lemma is only an unproved claim as far as Isabelle is concerned, so we now have to 
prove it. Proving in Isabelle rests on two pillars:
First, an intuitive, human-readable formal proof language, called \emph{Isar}, for proving theorems 
interactively. That means, the user writes down the individual steps of the proof, and Isabelle 
checks whether they are indeed correct. Second, a huge machinery of automatic proof methods that are able to prove certain goals 
automatically, saving the user from doing tedious but more or less simple proofs manually. 
Existing automation is fairly sophisticated, incorporating even powerful state-of-art 
first-order reasoners.

To give a rough idea of how proofs in Isabelle/HOL look like, we show a quite verbose induction
proof of the above lemma; long dashes ($\--$) indicate explanatory comments:
\begin{lstlisting}
lemma set-append: set (xs @ ys) = set xs $\cup$ set ys
proof (induction xs)
  $\--$Induction base:
  show set ([] @ ys) = set [] $\cup$ set ys by simp
	  $\--$Prove the goal %by% simplification w.r.t. the definitions of %\guilsinglleft%set%\guilsinglright% %and% %\guilsinglleft%@%\guilsinglright%.
next
  $\--$Induction step:
  fix x xs  	$\--$Choose fresh %\guilsinglleft%x%\guilsinglright% %and% %\guilsinglleft%xs%\guilsinglright% arbitrary, but fixed.
  assume set (xs @ ys) = set xs $\cup$ set ys  	$\--$Assume the induction hypothesis.
  then show set ((x # xs) @ ys) = set (x # xs) $\cup$ set ys by simp
	  $\--$Prove the goal again %by% simplification,
	    but this time also %using% the induction hypothesis.
qed
\end{lstlisting}
Since we will not present any Isabelle-proofs in the remainder of this paper, we do not say more 
about proving in Isabelle here.

Finally, please note that more complicated lemmas involving assumptions can be stated following the 
`fixes'/`assumes'/`shows' pattern, to increase readability:
\begin{lstlisting}
lemma times-mono-int:
  fixes a b c :: int
  assumes a $\leq$ b and 0 $\leq$ c
  shows a * c $\leq$ b * c
\end{lstlisting}
The optional `fixes' clause locally fixes variables and potentially annotates them with types 
(here \textsf{int}, the type of integers). The optional `assumes' clause states one or more 
assumptions, and the mandatory `shows' clause states the ultimate conclusion.

\begin{notation}
Within enclosing informal text, names of lemmas and theorems will be typeset in \textit{italics}, 
like \textit{set-append}.
\end{notation}

\subsection{Frequently Used Functions}
\label{sec::IsaFreq}

We conclude this section by listing built-in concepts we will use later on.
\begin{itemize}
 \item The usual logical connectives and quantifiers. Syntax in Isabelle/HOL closely 
resembles ordinary mathematical notation, expect that logical implication is denoted by 
$\longrightarrow$ and equivalence by $\longleftrightarrow$.

 \item The usual operations from set theory, whose Isabelle-syntax resembles mathematical notation, 
too.

 \item \textsf{$f$\ \textasciigrave $A$}, which denotes the image of set $A$ under function $f$.
 
 \item \textsf{\{0..<$n$\}}, which denotes the set $\{k\in\mathbb{N}\ |\ 0\leq k<n\}$ of natural numbers; analogously, \textsf{[0..<$n$]} denotes the list of natural numbers from $0$ up to 
$n$.

 \item \textsf{set}, \textsf{[]} and \textsf{\#}, as explained above: $\textsf{set}(xs)$ is 
the set of elements of list $xs$, \textsf{[]} is the empty list, and $x\,\#\,xs$ is the list whose 
first element is $x$ and whose tail is $xs$.

 \item $\textsf{length}(xs)$, which is the length of list $xs$.
 
 \item $xs\ !\ i$, which is the $i$-th element of list $xs$, starting from $0$.
 
 \item \textsf{fst} and \textsf{snd}, which project pairs of type $\alpha\times\beta$ onto their 
first and second entries, respectively. For instance, $\textsf{fst}((a,b))=a$.
\end{itemize}

\begin{rem}
Throughout the paper we will follow the common convention in papers about Isabelle of using dashes 
instead of underscores, for the sake of better readability. So, \textsf{subset-eq} would in reality 
be \textsf{subset\_eq} in the actual Isabelle sources.
\end{rem}

% 2 pages

\section{Preliminaries}
\label{sec::Preliminaries}

\subsection{Mathematical Preliminaries}
\label{sec::PreliminariesMath}

In this and the subsequent sections we present signature-based Gr\"obner basis algorithms and their 
formalization in Isabelle/HOL. Notation is mainly borrowed from~\cite{Eder2017}, with some small 
adjustments here and there to resemble the
notation we use in the formalization. In fact, since the formalization itself closely follows 
Sections~4--7 of~\cite{Eder2017}, most of the mathematical details omitted in this exposition for 
the sake of brevity can be found there instead. The informal proofs that served as the templates 
for our formal development were exclusively taken from the above-mentioned article 
and from~\cite{Roune2012} and~\cite{Eder2013}.

In the remainder of this paper let $\KK$ be a field and let $\RR=\KK[x_1,\ldots,x_n]$ be the 
$n$-variate polynomial ring over $\KK$. Every polynomial $p\in\RR$ can be written as a 
$\KK$-linear combination of \emph{power-products}, where a power-product is a product of the 
indeterminates $x_1,\ldots,x_n$, e.\,g.\ $x_1^2\,x_2\,x_3^4$. We will write $[X]$ for the commutative 
monoid of power-products in $x_1,\ldots,x_n$ and typically denote power-products by the typed 
variables $s$ and $t$, unless stated otherwise.

Now, fix a finite sequence $F=(f_1,\ldots,f_m)$ of polynomials in 
$\RR$; these polynomials play the role of the set we want to compute a Gr\"obner basis of. The sequence $F$ 
gives rise to a module-homomorphism $\overline{\cdot}:\RR^m\rightarrow\RR$ by setting 
$\overline{\mathbf{e}_i}:= f_i$ for $1\leq i\leq m$ and canonical basis vectors $\mathbf{e}_i$ 
of the free module $\RR^m$. A module element $a\in\RR^m$ is called a \emph{syzygy} of $F$ if 
$\overline{a}=0$. Note that $\RR^m$ can be viewed as a $\KK$-vector space, meaning that every $a\in 
\RR^m$ can be written as a $\KK$-linear combination of \emph{terms}, where a term is a product of 
the form $t\,\mathbf{e}_i$ for some power-product $t$ and some $1\leq i\leq m$. We will write $\TT$ 
for the set of terms and typically denote terms by the typed variables $u$ and $v$. For a term 
$v=t\,\mathbf{e}_i$, $t$ is called the power-product of $v$ and $i$ is called the component of $v$.

For a polynomial $p\in\RR$, $\supp(p)$ is the \emph{support} of $p$, which the set of all 
power-products appearing in $p$ with non-zero coefficient. Likewise, $\supp(a)$ for $a\in\RR^m$ is 
the set of all terms appearing in $a$ with non-zero coefficient. $\coeff(p,t)$ denotes the 
coefficient of power-product $t$ in $p\in\RR$, and analogous for $\coeff(a,u)$ with $u\in\TT$ and 
$a\in\RR^m$.

Finally we must also fix an admissible order relation $\preceq$ on $[X]$ and 
some compatible extension $\preceqt$ to a term order on $\TT$. \emph{Admissible} has the usual meaning of $1\preceq t$ and $s_1\preceq s_2\Rightarrow t\,s_1\preceq t\,s_2$ for all $s_1,s_2,t\in [X]$. \emph{Compatible} just means that $s\preceq t\Rightarrow s\,\mathbf{e}_i\preceqt t\,\mathbf{e}_i$
for all $s,t\in [X]$ and $1\leq i\leq m$. The most important extension of $\preceq$ to a term order 
is the \emph{position over term} (POT) extension, denoted by $\preceqpot$ and defined as
$$
s\,\mathbf{e}_i\preceqpot 
t\,\mathbf{e}_j :\Leftrightarrow i<j\vee (i=j\wedge s\preceq t).
$$

Every $p\in\RR$ has a \emph{leading power-product} $\lp(p)$ and a \emph{leading 
coefficient} $\lc(p)$: if $p\neq 0$, the leading power-product of $p$ is the largest power-product \wrt\ $\preceq$ appearing in $\supp(p)$, and the leading coefficient is its coefficient; 
$\lp(0)$ is left undefined and $\lc(0):=0$. 
Likewise, every module element $a\in\RR^m$ has a \emph{leading term} $\sig(a)$ and a 
leading coefficient $\lc(a)$, defined completely analogously \wrt\ $\preceqt$. The reason why the 
leading term of $a$ is denoted by $\sig(a)$ rather than $\mathrm{lt}(a)$ becomes clear in the 
following definition:

\begin{dfn}[Signature]
Let $a\in\RR^m$. The \emph{signature} of $a$ is the leading term $\sig(a)$ of $a$.
\end{dfn}
Therefore, the all-important \emph{signature} of a module element $a\in\RR^m$ is nothing else but 
the leading term of $a$. In the remainder, we will exclusively use the word `signature' instead of 
`leading term'. Note that in contrast to~\cite{Eder2017}, in our case the signature only consists 
of a term \emph{without} coefficient.

Summarizing, every $a\in\RR^m$ has two important values associated to it: its signature 
$\sig(a)\in\TT$ and the polynomial $\overline{a}\in\RR$.

\subsection{Isabelle/HOL}
\label{sec::PreliminariesIsabelle}

For our formal development of signature-based Gr\"obner basis algorithms we did not have to start 
completely from scratch, but could build upon on existing extensive formalizations of multivariate 
polynomials and Gr\"obner bases. Here, we will explain the most important aspects of these 
formalizations that are relevant for signature-based algorithms; 
the interested reader is referred to~\cite{Maletzky2018,Maletzky2018a} for more details. The 
formalizations are freely available as separate entries in the Archive of Formal Proofs 
(\cite{Sternagel2010,Immler2016}).

In Isabelle/HOL, multivariate polynomials are represented as so-called \emph{polynomial mappings}, 
which are functions from some type $\alpha$ to another type $\beta$ such that all but finitely many 
values are mapped to $0$. The meaning of such a mapping is clear: $\alpha$ plays the role of the 
power-products or terms, and $\beta$ plays the role of the coefficient-ring; the value a 
power-product or term is mapped to is its coefficient in the polynomial. The type of 
polynomial mappings from $\alpha$ to $\beta$ is denoted by \polymapping{\alpha}. Note that polynomial mappings are sometimes also referred to as \emph{finitely supported functions}.

Terms are simply represented as pairs consisting of a power-product and a component of type 
\textsf{nat}, the type of natural numbers; hence, if $\alpha$ is the type of power-products, then 
the type of terms is $\alpha\times\textsf{nat}$. Note that because of \textsf{nat} being infinite, 
the components of terms can become arbitrarily large; this point deserves a bit more attention, 
which will be paid below. Before, we summarize what we have so far:
\begin{itemize}
 \item Here and henceforth, $\alpha$ will always denote the type of power-products. How exactly 
power-products are represented is not important, since they essentially only have to form a 
cancellative commutative monoid and a lattice \wrt\ divisibility. For more details 
see~\cite{Maletzky2018a}.

 \item Type $\tau$ will abbreviate the type of terms, i.\,e., the type $\alpha\times\textsf{nat}$. In the 
actual formalization, $\tau$ only needs to be isomorphic to $\alpha\times\textsf{nat}$, but this is 
a mere technicality without any further implications.

 \item Type $\beta$ plays the role of $\KK$, the coefficient field.

 \item The polynomial ring $\RR$ hence corresponds to the type \polymapping{\alpha}, and the 
module $\RR^m$ to \polymapping{\tau}.
\end{itemize}

\begin{ex}
Let $p=3x^2-2xy+y^3-4\in\KK[x,y]$. Then $p$ corresponds to an object of type 
$\polymapping{\alpha}$ which maps $x^2\mapsto 3$, $xy\mapsto -2$, $y^3\mapsto 1$, $1\mapsto -4$, 
and all other power-products $t\mapsto 0$. As indicated above, how the individual power-products 
are represented is not important here.

Likewise, let
\[
a=\begin{pmatrix}2xy^2-3\\x^3+y^3\end{pmatrix}\in\KK[x,y]^2.
\]
Then $a$ corresponds to an object of type $\polymapping{\tau}$ which maps $(xy^2,0)\mapsto 2$, 
$(1,0)\mapsto -3$, $(x^3,1)\mapsto 1$, $(y^3,1)\mapsto 1$, and all other terms $u\mapsto 0$. Note 
in particular that the first component is indexed by $0$, the second by $1$, etc.
\end{ex}

One may now ask the following legitimate question:
\begin{quote}
How can $\RR^m$ be represented by 
$\polymapping{\tau}$, if $\RR^m$ has dimension $m$ (and therefore all terms appearing in its 
elements have components $\leq m$), but $\tau$ allows for arbitrarily large components?
\end{quote}
Indeed, strictly speaking $\RR^m$ and $\polymapping{\tau}$ are not isomorphic. However, we can 
circumvent the problem of components of terms being greater than $m$ (or, in fact $m-1$, since the 
first component is indexed by $0$) by explicitly putting certain constraints on all module elements 
of type $\polymapping{\tau}$ appearing anywhere in the formal development. More precisely, we 
define the set $\textsf{sig-inv-set}(m)$, parameterized over the natural number $m$, of all module 
elements whose terms have components in the range $[0,\ldots,m-1]$. Then, we constrain all theorems 
where it is necessary by the additional condition that all module elements $a$ occurring in the 
theorem belong to $\textsf{sig-inv-set}(m)$.

Of course, $m$ is not just an arbitrary natural number, but it is the length of the implicitly 
fixed sequence $F=(f_1,\ldots,f_m)$ of polynomials. So, in the formalization we also fix a 
sequence, or more precisely a list, \fs\ in the implicit theory context:
\begin{lstlisting}
context fixes fs :: ($\polymapping{\alpha}$) list
\end{lstlisting}
This instruction ensures that all subsequent definitions, theorems, algorithms, etc. are implicitly 
parameterized over the list \fs. Consequently, we can define the set \textsf{\RRm} of all  `valid' 
module elements:
\begin{lstlisting}
definition %\RRm% :: ($\polymapping{\tau}$) set
  where %\RRm% = sig-inv-set (length fs)
\end{lstlisting}
Hence, \textsf{\RRm} is the set of module elements whose terms have components in the range
$[0,\ldots,\textsf{length}(\fs)-1]$, and as such precisely corresponds to $\RR^m$.

\begin{rem}
In a dependently-typed system like Coq~(\cite{Bertot2004}), \textsf{\RRm} could be turned 
into a type, meaning that the additional assumptions $a\in\textsf{\RRm}$ of theorems could be 
encoded implicitly in the type of $a$. Isabelle/HOL is only simply-typed, so this approach does not 
work in our case.
\end{rem}

\begin{rem}
The definition of \textsf{\RRm} shown above is not exactly the one of the formalization. 
Namely, analogous to components, we also have to take care that all indeterminates appearing in 
module elements also appear in \fs; similar as for the components, this cannot be encoded (easily) 
in the type $\alpha$. The details are technical and omitted here for the sake of simplicity.
\end{rem}

Having \fs\ fixed in the context, we next formalize the module-homomorphism $\overline{\cdot}$, 
called \textsf{\poly} in the formal theory, and prove its characteristic properties. We omit its slightly technical definition here.
\begin{lstlisting}
lemma %\poly%-zero: %\poly% 0 = 0%\\[-1ex]%
lemma %\poly%-plus: %\poly% (a + b) = %\poly% a + %\poly% b%\\[-1ex]%
lemma %\poly%-mult-scalar: %\poly% (p $\odot$ a) = p * %\poly% a
\end{lstlisting}
Here one should note that \textsf{poly} is defined in such a way that these identities hold 
unconditionally, even if $a,b\notin\textsf{\RRm}$. The expression $p\odot a$ denotes scalar multiplication of the 
module element $a$ by the polynomial $p$. Two further important lemmas about \textsf{\poly} 
describe its relationship to the ideal generated by the elements of \fs:
\begin{lstlisting}
lemma %\poly%-in-ideal: %\poly% a $\in$ ideal (set fs)%\\[-1ex]%
lemma in-idealE-%\poly%-%\RRm%:
  assumes p $\in$ ideal (set fs)
  shows $\exists$a$\in$%\RRm%. p = %\poly% a
\end{lstlisting}
The first lemma obviously expresses that $\textsf{\poly}(a)$ is always an element of the ideal 
generated by the elements of \fs, whereas the second lemma states the converse: every element $p$ 
of the ideal can be written as $p=\textsf{\poly}(a)$ for some $a\in\textsf{\RRm}$; $a$ being an 
element of \textsf{\RRm} is of particular importance here. The expression $\textsf{ideal}(B)$, taken from~\cite{Sternagel2010}, denotes the ideal generated by the set $B$.

The only things that are still missing from Section~\ref{sec::PreliminariesMath} are the order 
relations $\preceq$ and $\preceqt$, and the various concepts they induce (leading power-product 
etc.); they are, in fact, contained in~\cite{Sternagel2010} as well. Similar to \fs, the remaining formal development shall be parameterized over these 
orderings, so a \emph{locale} is employed to fix them implicitly:
\begin{lstlisting}
locale qpm-inf-term =
  ordered-powerprod ord +
  linorder ord-term
    for ord :: $\alpha$ $\Rightarrow$ $\alpha$ $\Rightarrow$ bool (infixl $\preceq$ 50)
    and ord-term :: $\tau$ $\Rightarrow$ $\tau$ $\Rightarrow$ bool (infixl $\preceqt$ 50) +
  assumes stimes-mono: u $\preceqt$ v $\longrightarrow$ t $\otimes$ u $\preceqt$ t $\otimes$ v
  assumes ord-termI: fst u $\preceq$ fst v $\longrightarrow$ snd u $\leq$ snd v $\longrightarrow$ u $\preceqt$ v
\end{lstlisting}
Locales are a sophisticated mechanism for structuring Isabelle-theories into sub-theories that can 
later be combined in a convenient and efficient way. Before, when fixing \fs, a simple \textsf{context}-statement was sufficient, but here we really need the capabilities of a full-fledged locale. For more information on locales see~\cite{Ballarin2010}. 

Locale \textsf{qpm-inf-term} fixes the two relations $\preceq$ and $\preceqt$, and assumes that $\preceq$ is an 
admissible order on type $\alpha$ (the power-products) and that $\preceqt$ is a linear order on 
type $\tau$ (the terms). Furthermore, it assumes the two properties \textit{stimes-mono} and \textit{ord-termI}. Property \textit{stimes-mono} expresses that $\preceqt$ is monotonic \wrt\ $\otimes$, where $t\otimes u$ 
denotes the term obtained from $u$ by multiplying its power-product by $t$. Property \textit{ord-termI} 
states that if the power-product and the component of $u$ are not greater than their respective 
counterparts of $v$, then $u\preceqt v$. Note that $\textsf{fst}(u)$ gives the first entry of term 
$u$, i.\,e.\ its power-product, and $\textsf{snd}(u)$ gives the second entry of $u$, i.\,e.\ its 
component.

All subsequent definitions, theorems, etc. will be stated in the context of this locale, meaning 
that they are implicitly parameterized over $\preceq$ and $\preceqt$ (just as they are 
parameterized over \fs), and that furthermore all theorems are implicitly constrained by the two 
additional assumptions \textit{stimes-mono} and \textit{ord-termI}. Leading power-products, leading terms and signatures of polynomials and module elements can 
be defined readily in this setup.

We conclude this section by pointing the reader to \ref{apx::Translation}, containing a 
glossary for translating between mathematical notions and notations occurring in this paper, and 
their counterparts in the formalization.

\begin{rem}
In the actual Isabelle sources, power-products are written \emph{additively} rather than \emph{multiplicatively}, for technical 
reasons. So, $0$ is used instead of $1$, $+$ instead of $\cdot$, and $\oplus$ instead of $\otimes$. 
In this paper we decided to stick to the standard multiplicative writing for the sake of uniformity.
\end{rem}

% 4.5 pages

\section{Signature Reduction and Signature Gr\"obner Bases}
\label{sec::Reduction}

Let us now turn to the key concept in the theory of Gr\"obner bases: \emph{polynomial reduction}. 
In the `traditional', non-signature approach the reduction relation is a binary relation on 
polynomials, parameterized over a set of polynomials. In the signature-based world, it becomes a 
binary relation on \emph{module elements}, i.\,e.\ on $\RR^m$, defined as follows:
\begin{dfn}[\sig-Reduction]
\label{dfn::Reduction}
Let $a,b\in\RR^m$ and $G\subseteq\RR^m$. The module element $a$ \sig-reduces to $b$ modulo $G$ if, and only if, there 
exist $g\in G$ and $t\in [X]$ such that
\begin{enumerate}
 \item $\overline{g}\neq 0$,
 \item $t\,\lp(\overline{g})\in\supp(\overline{a})$,
  \item $b=a-t\,g$, and
 \item $t\,\sig(g)\preceqt\sig(a)$, which is equivalent to $\sig(b)\preceqt\sig(a)$.
\end{enumerate}
\end{dfn}
So, if $a$ \sig-reduces to $b$ modulo $G$, it simply means that $\overline{a}$ reduces to 
$\overline{b}$ modulo $\overline{G}$ in the usual sense of polynomial 
reduction,\footnote{$\overline{G}$, of course, denotes the image of $G$ under the homomorphism 
$\overline{\cdot}$.} and that furthermore the signature of $b$ is not greater than that of 
$a$. In short, \sig-reduction is like polynomial reduction with the additional requirement that 
signatures do not grow.

\sig-reduction comes in different flavors, depending on whether $t\,\lp(\overline{g})$ in 
Definition~\ref{dfn::Reduction} equals $\lp(\overline{a})$ or not, and whether in the last 
condition of that definition we have $t\,\sig(g)=\sig(a)$ or $t\,\sig(g)\prect\sig(a)$. If 
$t\,\lp(\overline{g})=\lp(\overline{a})$, we shall say that the \sig-reduction is a \emph{top}
\sig-reduction; otherwise, if $t\,\lp(\overline{g})\prec\lp(\overline{a})$, we call it a \emph{tail} 
\sig-reduction.\footnote{Obviously $\lp(\overline{a})\prec t\,\lp(\overline{g})$ is not possible,
since $t\,\lp(\overline{g})\in\supp(\overline{a})$.} If $t\,\sig(g)=\sig(a)$ the \sig-reduction is a
\emph{singular} \sig-reduction, whereas if $t\,\sig(g)\prect\sig(a)$ it is a \emph{regular}
\sig-reduction. It is easy to see that in a regular \sig-reduction we always have $\sig(b)=\sig(a)$.

\begin{notation}
\label{ntn::Reduction}
Let $\mathrm{r}_1\in\{\preceqt,\prect,=\}$ and $\mathrm{r}_2\in\{\preceq,\prec,=\}$. We will use 
the following notation: $a\red{\mathrm{r}_1}{\mathrm{r}_2}{G} b$ means that $a$ \sig-reduces to $b$ 
modulo $G$, and that additionally $t\,\sig(g)\ \mathrm{r}_1\ \sig(a)$ and
$t\,\lp(\overline{g})\ \mathrm{r}_2\ \lp(\overline{a})$ hold, where $t,g$ are as in
Definition~\ref{dfn::Reduction}. As usual, $\red{\mathrm{r}_1}{\mathrm{r}_2}{G}^*$ denotes the
reflexive-transitive closure of $\red{\mathrm{r}_1}{\mathrm{r}_2}{G}$. So,
$\red{\preceq}{\preceqt}{G}$ stands for general \sig-reduction, $\red{\prect}{=}{G}$ for regular top
\sig-reduction, and so on. To ease notation, we will simply write $\longrightarrow_G$ instead of
$\red{\preceqt}{\preceq}{G}$.
\end{notation}

What has been said above about the relationship between $\sig(b)$ and $\sig(a)$ if $b$ \sig-reduces 
to $a$ is of course also true for the reflexive-transitive closure of \sig-reduction. In 
particular, if $a\red{\prect}{\mathrm{r}_2}{G}^* b$, i.\,e.\ $a$ regular \sig-reduces to $b$ in 
several steps, then $\sig(b)=\sig(a)$. This trivial observation will play a crucial role later on.

The definition of \sig-reduction in the formalization closely follows 
Definition~\ref{dfn::Reduction}, but in addition also incorporates Notation~\ref{ntn::Reduction}:
\begin{lstlisting}
definition sig-red-single :: ($\tau$ $\Rightarrow$ $\tau$ $\Rightarrow$ bool) $\Rightarrow$ ($\alpha$ $\Rightarrow$ $\alpha$ $\Rightarrow$ bool) $\Rightarrow$
			      ($\polymapping{\tau}$) $\Rightarrow$ ($\polymapping{\tau}$) $\Rightarrow$ ($\polymapping{\tau}$) $\Rightarrow$ $\alpha$ $\Rightarrow$ bool
  where sig-red-single r1 r2 a b g t $\longleftrightarrow$
	  (%\poly% g $\neq$ 0 $\wedge$ coeff (%\poly% a) (t * lp (%\poly% g)) $\neq$ 0 $\wedge$
	    b = a - monom-mult ((coeff (%\poly% a) (t * lp (%\poly% g))) / lc (%\poly% g)) t g $\wedge$
	    r1 (t $\otimes$ $\sig$ g) ($\sig$ a) $\wedge$ r2 (t * lp (%\poly% g)) (lp (%\poly% a)))%\\[-1ex]%
definition sig-red :: ($\tau$ $\Rightarrow$ $\tau$ $\Rightarrow$ bool) $\Rightarrow$ ($\alpha$ $\Rightarrow$ $\alpha$ $\Rightarrow$ bool) $\Rightarrow$
		       ($\polymapping{\tau}$) set $\Rightarrow$ ($\polymapping{\tau}$) $\Rightarrow$ ($\polymapping{\tau}$) $\Rightarrow$ bool
  where sig-red r1 r2 G a b $\longleftrightarrow$ ($\exists$g$\in$G. $\exists$t. sig-red-single r1 r2 a b g t)%\\[-1ex]%
definition is-sig-red :: ($\tau$ $\Rightarrow$ $\tau$ $\Rightarrow$ bool) $\Rightarrow$ ($\alpha$ $\Rightarrow$ $\alpha$ $\Rightarrow$ bool) $\Rightarrow$
			  ($\polymapping{\tau}$) set $\Rightarrow$ ($\polymapping{\tau}$) $\Rightarrow$ bool
  where is-sig-red r1 r2 G a $\longleftrightarrow$ ($\exists$b. sig-red r1 r2 G a b)
\end{lstlisting}
So, $\textsf{sig-red-single}(\mathrm{r}_1,\mathrm{r}_2,a,b,g,t)$ expresses that $a$ \sig-reduces to 
$b$ modulo the singleton $\{g\}$ using the given power-product $t$ as the 
multiplier. Note that $\textsf{monom-mult}(c,t,a)$ is multiplication of $a$ by the coefficient $c$ 
and power-product $t$. The two relations $\mathrm{r}_1$ and $\mathrm{r}_2$ have exactly the same 
meaning as in Notation~\ref{ntn::Reduction}, i.\,e., they specify whether the \sig-reduction is 
singular/regular/arbitrary and top/tail/arbitrary, respectively. 
The expression $\textsf{sig-red}(\mathrm{r}_1,\mathrm{r}_2,G,a,b)$ precisely corresponds to 
$a\red{\mathrm{r}_1}{\mathrm{r}_2}{G}b$; the reflexive-transitive closure of \sig-reduction is thus
given by $\textsf{sig-red}(\mathrm{r}_1,\mathrm{r_2},G)^{**}$, using Isabelle/HOL's built-in 
notation $r^{**}$ for denoting the reflexive-transitive closre of an arbitrary binary relation $r$. 
Finally, \textsf{is-sig-red} is an auxiliary notion expressing \sig-reducibility.

Since traditional polynomial reduction is known to be Noetherian, and \sig-reduction in 
some sense `refines' polynomial reduction, we can immediately infer that \sig-reduction is 
Noetherian, too:
\begin{lem}
For all $G\subseteq\RR^m$, $\red{\mathrm{r}_1}{\mathrm{r}_2}{G}$ is Noetherian, that is, there are 
no infinite chains $a_1\red{\mathrm{r}_1}{\mathrm{r}_2}{G} a_2\red{\mathrm{r}_1}{\mathrm{r}_2}{G} 
\ldots$.
\end{lem}
This lemma can be translated easily into Isabelle/HOL, employing the built-in predicate 
\textsf{wfP} for expressing well-foundedness of the converse of \sig-reduction (denoted by $^{--}$):
\begin{lstlisting}
lemma sig-red-wf-%\RRm%:
  assumes G $\subseteq$ %\RRm%
  shows wfP (sig-red r1 r2 G)$^{--}$
\end{lstlisting}
Proving this lemma is a matter of only a couple of lines, thanks to the fact that 
\cite{Immler2016} already proved Noetherianity of traditional polynomial reduction in Isabelle/HOL. 
The assumption of $G$ being a subset of \textsf{\RRm} is necessary because of the observations made 
in Section~\ref{sec::PreliminariesIsabelle}.

Before we define signature Gr\"obner bases, we introduce an auxiliary notion:
\begin{dfn}
\label{dfn::RedZero}
We say that $a$ \emph{\sig-reduces to zero} modulo $G$ if, and only if, there exists $b$ such that 
$a\red{\preceqt}{\preceq}{G}^*b$ and $\overline{b}=0$, i.\,e., $b$ is a syzygy. 
Just as for \sig-reduction, we will also use the phrases \emph{singular} and \emph{regular} 
\sig-reduction to zero, if $a\red{=}{\preceq}{G}^*b$ or $a\red{\prect}{\preceq}{G}^*b$, 
respectively.
\end{dfn}
Note that even though we use the word `zero' in Definition~\ref{dfn::RedZero} it does not mean that 
$b$ itself has to be $0$, only that it must be a syzygy. This terminology is taken 
from~\cite{Eder2017}.

In the non-signature world, a Gr\"obner basis is a set $G\subseteq\RR$ such that every $p\in\langle 
G\rangle$ can be reduced to $0$ modulo $G$. This definition can be translated readily into the 
signature-based setting:
\begin{dfn}[Signature Gr\"obner Basis]
\label{dfn::SigGB}
Let $u$ be a term. A set $G\subseteq\RR^m$ is a \emph{signature Gr\"obner basis} in $u$ if, and only 
if, every $a\in\RR^m$ with $\sig(a)=u$ \sig-reduces to zero modulo $G$.

The set $G$ is a signature Gr\"obner basis up to $u$ if it is a signature Gr\"obner basis in all $v\prect 
u$. If $G$ is a signature Gr\"obner basis in all terms, we simply call it a signature Gr\"obner 
basis.
\end{dfn}
Translating Definitions~\ref{dfn::RedZero} and~\ref{dfn::SigGB} into Isabelle/HOL is again 
immediate; the only real differences are some \textsf{\RRm} conditions, as usual:
\begin{lstlisting}
definition sig-red-zero :: ($\tau$ $\Rightarrow$ $\tau$ $\Rightarrow$ bool) $\Rightarrow$ ($\polymapping{\tau}$) set $\Rightarrow$ ($\polymapping{\tau}$) $\Rightarrow$ bool
  where sig-red-zero r1 G a $\longleftrightarrow$ ($\exists$b. (sig-red r1 ($\preceq$) G)$^{**}$ a b $\wedge$ %\poly% b = 0)%\\[-1ex]%
definition is-sig-GB-in :: ($\polymapping{\tau}$) set $\Rightarrow$ $\tau$ $\Rightarrow$ bool
  where is-sig-GB-in G u $\longleftrightarrow$
		  ($\forall$a. $\sig$ a = u $\longrightarrow$ a $\in$ %\RRm% $\longrightarrow$ sig-red-zero ($\preceqt$) G a)%\\[-1ex]%
definition is-sig-GB_upt :: ($\polymapping{\tau}$) set $\Rightarrow$ $\tau$ $\Rightarrow$ bool
  where is-sig-GB-upt G u $\longleftrightarrow$
	(G $\subseteq$ %\RRm% $\wedge$ ($\forall$v. v $\prect$ u $\longrightarrow$ snd v < length fs $\longrightarrow$ is-sig-GB-in G v))
\end{lstlisting}
Please note that \textsf{sig-red-zero} is only parameterized over the relation $\mathrm{r}_1$ for
signatures, but not over $\mathrm{r}_2$ for leading power-products: there is no need to distinguish
between top/tail/arbitrary \sig-reductions to zero.

The connection between signature Gr\"obner bases and ordinary non-signature Gr\"obner bases follows 
immediately from the definition of \sig-reduction:
\begin{prop}
\label{prop::SigGB}
Let $G\subseteq\RR^m$ be a signature Gr\"obner basis. Then $\overline{G}$ is a 
Gr\"obner basis of $\langle f_1,\ldots,f_m\rangle$.
\end{prop}

Signature-based Gr\"obner basis algorithms, such as $F_5$, compute \emph{rewrite bases}, which are a subclass of signature Gr\"obner bases (see Section~\ref{sec::RewriteBases}).
Proposition~\ref{prop::SigGB} tells us that from a signature Gr\"obner basis one can easily obtain 
a Gr\"obner basis of the ideal $\langle f_1,\ldots,f_m\rangle$ under consideration by applying the 
module-homomorphism $\overline{\cdot}$ to all elements.

The formalization of Proposition~\ref{prop::SigGB} looks as
follows, where \textsf{is-Groebner-basis} is defined in~\cite{Immler2016}:
\begin{lstlisting}
lemma is-sig-GB-is-Groebner-basis:
  assumes G $\subseteq$ %\RRm% and $\forall$u.%\,%is-sig-GB-in G u
  shows is-Groebner-basis (%\poly% %\textasciigrave% G)
\end{lstlisting}

The next result about signature Gr\"obner bases will prove very useful later on, for instance in 
Lemma~\ref{lem::SyzCrit}. Readers interested in its proof are referred to Lemma~3 
in~\cite{Roune2012}.
\begin{lem}
\label{lem::RedUnique}
Let $a,b\in\RR^m\backslash\{0\}$, let $G$ be a signature Gr\"obner basis up to $\sig(a)$, and assume
$\sig(a)=\sig(b)$ and $\lc(a)=\lc(b)$.
\begin{enumerate}
 \item If both $a$ and $b$ are regular top \sig-irreducible modulo $G$, then 
$\lp(\overline{a})=\lp(\overline{b})$ and $\lc(\overline{a})=\lc(\overline{b})$.

 \item If both $a$ and $b$ are regular \sig-irreducible modulo $G$, then 
$\overline{a}=\overline{b}$.
\end{enumerate}
\end{lem}
In the formalization this lemma is split into two lemmas:
\begin{lstlisting}
lemma sig-regular-top-reduced-lp-lc-unique:
  assumes is-sig-GB-upt G ($\sig$ a) and a $\in$ %\RRm% and b $\in$ %\RRm%
    and $\sig$ a = $\sig$ b and lc a = lc b
    and $\neg$ is-sig-red ($\prect$) (=) G a and $\neg$ is-sig-red ($\prect$) (=) G b
  shows lp (%\poly% a) = lp (%\poly% b) and lc (%\poly% a) = lc (%\poly% b)%\\[-1ex]%
lemma sig-regular-reduced-unique:
  assumes is-sig-GB-upt G ($\sig$ a) and a $\in$ %\RRm% and b $\in$ %\RRm%
    and $\sig$ a = $\sig$ b and lc a = lc b
    and $\neg$ is-sig-red ($\prect$) ($\preceq$) G a and $\neg$ is-sig-red ($\prect$) ($\preceq$) G b
  shows %\poly% a = %\poly% b
\end{lstlisting}

We conclude this section by introducing the concept of a \emph{syzygy signature} and proving an
important lemma about it:
\begin{dfn}[Syzygy Signature]
\label{dfn::SyzSig}
A term $u$ is called a \emph{syzygy signature} if there exists 
$a\in\RR^m\backslash\{0\}$ with $\sig(a)=u$ and $\overline{a}=0$.
\end{dfn}
Syzygy signatures play a key role for detecting useless zero-reductions when computing signature 
Gr\"obner bases. Namely, by virtue of Lemma~\ref{lem::RedUnique}, we obtain the following result 
whose importance will become clear in Section~\ref{sec::RewriteBases}:
\begin{lem}[Syzygy Criterion]
\label{lem::SyzCrit}
Let $a\in\RR^m$ and let $G$ be a signature Gr\"obner basis up to $\sig(a)$. If $\sig(a)$ is a syzygy
signature, then $a$ regular \sig-reduces to zero modulo $G$.

Moreover, if $u$ is a syzygy signature and $u\,|\,v$,\footnote{$u\,|\,v$, for two terms $u$ and 
$v$, means that there exists $t\in [X]$ with $v=t\,u$.} then $v$ is a syzygy signature, too.
\end{lem}

Definition~\ref{dfn::SyzSig} and Lemma~\ref{lem::SyzCrit} naturally
translate
into Isabelle/HOL:
\begin{lstlisting}
definition is-syz-sig :: $\tau$ $\Rightarrow$ bool
  where is-syz-sig u $\longleftrightarrow$ ($\exists$a$\in$%\RRm%. a $\neq$ 0 $\wedge$ $\sig$ a = u $\wedge$ %\poly% a = 0)%\\[-1ex]%
lemma syzygy-crit:
  assumes is-sig-GB-upt G ($\sig$ a) and is-syz-sig G ($\sig$ a) and a $\in$ %\RRm%
  shows sig-red-zero ($\prect$) G a%\\[-1ex]%
lemma is-syz-sig-dvd:
  assumes is-syz-sig u and u %\dvdt% v
  shows is-syz-sig v
\end{lstlisting}

% 4 pages

\section{Rewrite Bases and S-Pairs}
\label{sec::RewriteBases}

Besides signature Gr\"obner bases, we need another class of sets $G\subseteq\RR^m$ of module 
elements, called \emph{rewrite bases}. Rewrite bases play a crucial role for computing signature Gr\"obner bases, as will be seen in Section~\ref{sec::Algorithms}. Before, however, we must introduce some auxiliary concepts: sig-poly-pairs, rewrite orders and canonical rewriters.

%In order to define them, however, we first need to introduce the notion of a \emph{rewrite order}. We are aware that this notion comes `out of the blue' here, but think it is not in the scope of this exposition to give a detailed account on the motivation for and intuition behind rewrite orders. As usual, readers are referred to~\cite{Eder2017} instead.

\begin{dfn}[Sig-Poly-Pair]
\label{dfn::SPP}
A \emph{sig-poly-pair} is a pair $(u,p)\in\TT\times\RR$ such that there exists 
$a\in\RR^m\backslash\{0\}$ with $\sig(a)=u$ and $\overline{a}=p$.
\end{dfn}

Our definition of rewrite orders is slightly more technical than the one given in the literature. The 
reason for this deviation is that there, rewrite orders are defined for module elements rather than 
sig-poly-pairs. We found it more reasonable to define rewrite orders on sig-poly-pairs, 
because in any case the only information concrete rewrite orders may take into account for 
deciding which of the two arguments is greater are the signatures and the polynomial parts of the 
arguments.
\begin{dfn}[Rewrite Order]
\label{dfn::RewriteOrder}
A binary relation $\rw$ on sig-poly-pairs is called a \emph{rewrite order} if, and only 
if, it is a reflexive, transitive and linear relation, additionally satisfying
\begin{enumerate}
 \item $(u,p)\rw (v,q)\wedge (v,q)\rw (u,p)\Longrightarrow u=v$ for all 
sig-poly-pairs $(u,p)$ and $(v,q)$, and
 \item $a\in G\backslash\{0\}\wedge b\in G\backslash\{0\}\wedge \sig(a)\,|\,\sig(b)\Longrightarrow 
(\sig(a),\overline{a})\rw (\sig(b),\overline{b})$ for all $a,b,G$ such that $G$ is a 
signature Gr\"obner basis up to $\sig(b)$ and $b$ is regular top \sig-irreducible modulo $G$.
\end{enumerate}
\end{dfn}
The last condition essentially expresses that $\rw$ shall refine the divisibility relation on 
signatures---but only under some technical assumptions which are necessary for proving that 
$\rwrat$ (Definition~\ref{dfn::rwrat_rwadd}) is indeed a rewrite order.

To get some intuition about rewrite orders, we present the two `standard' rewrite orders 
that can be found in the literature:
\begin{dfn}[$\rwrat$, $\rwadd$]
\label{dfn::rwrat_rwadd}
The relation $\rwrat$ is defined as
\[
(u,p)\rwrat 
(v,q)\ :\Leftrightarrow\ \lp(q)\,u\prect\lp(p)\,v\vee(\lp(q)\,u=\lp(p)\,v\wedge u\preceqt v).
\]
The relation $\rwadd$ is defined as
\[
(u,p)\rwadd (v,q)\ :\Leftrightarrow\ u\preceqt v.
\]
\end{dfn}
As explained in Remark~7.3 in~\cite{Eder2017}, the suffix `\textsf{rat}' of $\rwrat$ originates 
from an alternative presentation of this relation, in which the \emph{ratios} 
$\frac{u}{\lp(p)}$ and $\frac{v}{\lp(q)}$ are compared.

Above we claimed that $\rwrat$ and $\rwadd$ are rewrite orders. The proof for $\rwadd$ is fairly 
straightforward, but the proof of the last requirement of rewrite orders is a bit more involved 
for $\rwrat$; one essentially has to make use of Lemma~\ref{lem::RedUnique} again.

The definition of rewrite orders in the formalization closely resembles 
Definition~\ref{dfn::RewriteOrder}; only note that $\textsf{spp-of}(a)$ is a mere abbreviation for 
$(\sig(a),\textsf{\poly}(a))$:
\begin{lstlisting}
definition is-rewrite-ord :: (($\tau$ $\times$ ($\polymapping{\alpha}$)) $\Rightarrow$ ($\tau$ $\times$ ($\polymapping{\alpha}$)) $\Rightarrow$ bool) $\Rightarrow$ bool
  where is-rewrite-ord ord $\longleftrightarrow$
		    (reflp ord $\wedge$ transp ord $\wedge$ ($\forall$a b. ord a b $\vee$ ord b a) $\wedge$
		     ($\forall$a b. ord a b $\longrightarrow$ ord b a $\longrightarrow$ fst a = fst b) $\wedge$
		     ($\forall$G a b. is-sig-GB-upt G ($\sig$ b) $\longrightarrow$ a $\in$ G $\longrightarrow$ b $\in$ G $\longrightarrow$
		       a $\neq$ 0 $\longrightarrow$ b $\neq$ 0 $\longrightarrow$ $\sig$ a $\dvdt$ $\sig$ b $\longrightarrow$
		       $\neg$ is-sig-red ($\prect$) (=) G b $\longrightarrow$ ord (spp-of a) (spp-of b)))
\end{lstlisting}
Since there is nothing special about the formal definitions of $\rwrat$ and $\rwadd$ compared to 
the informal ones, we omit them here.

Just as we have implicitly fixed $\preceq$ and $\preceqt$, let us now also fix an arbitrary rewrite 
order $\rw$. The last prerequisite we need before we can define rewrite bases are \emph{canonical 
rewriters}:
\begin{dfn}[Canonical Rewriter]
Let $G\subseteq\RR^m$, $a\in\RR^m$ and $u\in\TT$. The module element $a$ is called a \emph{canonical rewriter} in 
signature $u$ \wrt\ $G$ if, and only if, $a\in G\backslash\{0\}$, $\sig(a)\,|\,u$, and $a$ is 
maximal \wrt\ \rw\ with these properties.\footnote{By abuse of notation we also compare module 
elements in $\RR^m$ \wrt\ \rw, in the sense that $a\rw b\Leftrightarrow 
(\sig(a),\overline{a})\rw(\sig(b),\overline{b})$.}
\end{dfn}
\begin{dfn}[Rewrite Basis]
\label{dfn::RewriteBasis}
Let $G\subseteq\RR^m$ and $u\in\TT$. The set $G$ is said to be a \emph{rewrite basis} in $u$ if, and only 
if, $u$ is a syzygy signature or there exists a canonical rewriter $g$ in signature $u$ \wrt\ $G$ 
such that $\frac{u}{\sig(g)}g$ is regular top \sig-irreducible modulo $G$.\footnote{For two terms 
$u,v$ with $u\,|\,v$, $\frac{u}{v}$ denotes the unique $t\in [X]$ with $v=t\,u$.}

Furthermore, $G$ is a rewrite basis up to $u$ if it is a rewrite basis in all $v\prect u$. If $G$ is a rewrite 
basis in all terms, we simply call it a rewrite basis.
\end{dfn}

The definitions of canonical rewriters and rewrite bases translate naturally into in Isabelle/HOL; note in particular the parallels between the definitions of
\textsf{is-sig-GB-upt} and \textsf{is-RB-upt}:
\begin{lstlisting}
definition is-canon-rewriter :: ($\polymapping{\tau}$) set $\Rightarrow$ $\tau$ $\Rightarrow$ ($\polymapping{\tau}$) $\Rightarrow$ bool
  where is-canon-rewriter G u a $\longleftrightarrow$
		      (a $\in$ G $\wedge$ a $\neq$ 0 $\wedge$ $\sig$ a $\dvdt$ u $\wedge$
		       ($\forall$g$\in$G. g $\neq$ 0 $\longrightarrow$ $\sig$ g $\dvdt$ u $\longrightarrow$ spp-of g $\rw$ spp-of a)))%\\[-1ex]%
definition is-RB-in :: ($\polymapping{\tau}$) set $\Rightarrow$ $\tau$ $\Rightarrow$ bool
  where is-RB-in G u $\longleftrightarrow$
		(is-syz-sig u $\vee$
		 ($\exists$g. is-canon-rewriter G u g $\wedge$
		      $\neg$ is-sig-red ($\prect$) (=) G (monom-mult 1 (u / $\sig$ g) g)))%\\[-1ex]%
definition is-RB-upt :: ($\polymapping{\tau}$) set $\Rightarrow$ $\tau$ $\Rightarrow$ bool
  where is-RB-upt G u $\longleftrightarrow$
	    (G $\subseteq$ %\RRm% $\wedge$ ($\forall$v. v $\prect$ u $\longrightarrow$ snd v < length fs $\longrightarrow$ is-RB-in G v))
\end{lstlisting}

Now that we know what rewrite bases are it is time to establish the connection 
between rewrite bases and signature Gr\"obner bases, and hence to traditional Gr\"obner bases by 
virtue of Proposition~\ref{prop::SigGB}. For an informal proof of the following proposition, see 
Lemma~8 in~\cite{Eder2013}:
\begin{prop}
\label{prop::RB}
If $G$ is a rewrite basis up to $u$, it is also a signature Gr\"obner basis up to $u$.
\end{prop}
We omit the obvious translation of this proposition into Isabelle/HOL. Summarizing, 
Propositions~\ref{prop::SigGB} and~\ref{prop::RB} justify computing a rewrite basis in order to 
find a traditional Gr\"obner basis of $\langle f_1,\ldots,f_m\rangle$. Therefore, we now need a 
means for actually \emph{computing} rewrite bases---and it turns out that the key to an effective 
algorithm lies in a concept well-known from traditional Gr\"obner bases theory:
\begin{dfn}[S-Pair]
\label{dfn::S-pair}
Let $a,b\in\RR^m$, and let $t=\lcm(\lp(\overline{a}),\lp(\overline{b}))$. Then the \emph{S-pair} of 
$a$ and $b$, written $\spair(a,b)$, is defined as
\[
\spair(a,b)\ :=\ \frac{t}{\lc(\overline{a})\lp(\overline{a})}a-
\frac{t}{\lc(\overline{b})\lp(\overline{b})}b.
\]
Furthermore, $a$ and $b$ are said to give rise to a \emph{regular} S-pair if, and only if, 
$\overline{a},\overline{b}\neq 0$ and $\frac{t}{\lp(\overline{a})}\sig(a)\neq 
\frac{t}{\lp(\overline{b})}\sig(b)$; otherwise they give rise to a \emph{singular} S-pair.
\end{dfn}
So, S-pairs correspond precisely to S-polynomials, but `lifted' from $\RR$ to $\RR^m$: indeed, 
$\spair(a,b)\in\RR^m$, and it is easy to see that 
$\overline{\spair(a,b)}=\mathrm{spoly}(\overline{a},\overline{b})$, where $\mathrm{spoly}(p,q)$ 
is the usual S-polynomial of $p$ and $q$.

The distinction between singular and regular S-pairs is important, because in 
Theorem~\ref{thm::Main} below we will show that only regular S-pairs are of interest. If $a$ and 
$b$ give rise to a regular S-pair, we have 
$\sig(\spair(a,b))=\max(\frac{t}{\lp(\overline{a})}\sig(a),\frac{t}{\lp(\overline{b})}
\sig(b))$, where $t$ is as in Definition~\ref{dfn::S-pair}.

The definitions of S-pairs and regular S-pairs in the formalization look as follows:
\begin{lstlisting}
definition spair :: ($\polymapping{\tau}$) $\Rightarrow$ ($\polymapping{\tau}$) $\Rightarrow$ ($\polymapping{\tau}$)
  where spair a b = (let t1 = lp (%\poly% a); t2 = lp (%\poly% b); t = lcm t1 t2 in
                        (monom-mult (1 / lc (%\poly% a)) (t / t1) a) -
                        (monom-mult (1 / lc (%\poly% b)) (t / t2) b))%\\[-1ex]%
definition is-regular-spair :: ($\polymapping{\tau}$) $\Rightarrow$ ($\polymapping{\tau}$) $\Rightarrow$ bool
  where is-regular-spair a b $\longleftrightarrow$
                    (%\poly% a $\neq$ 0 $\wedge$ %\poly% b $\neq$ 0 $\wedge$
                     (let t1 = lp (%\poly% a); t2 = lp (%\poly% b); t = lcm t1 t2 in
                        (t / t1) $\otimes$ $\sig$ a $\neq$ (t / t2) $\otimes$ $\sig$ b))
\end{lstlisting}

Now we are ready to state the central theorem in this section, which links rewrite bases to 
regular S-pairs just as Buchberger's theorem links Gr\"obner bases to S-polynomials:
\begin{thm}
\label{thm::Main}
Let $G\subseteq\RR^m$ be finite and $u\in\TT$, assume that no two elements of $G$ have the same
signatures, and moreover assume that $G$ is a rewrite basis in all $\sig(a)\prect 
u$, where $a$ is either a regular S-pair of elements of $G$ or $a=\mathbf{e}_i$ ($1\leq i\leq m$). 
Then $G$ is a rewrite basis up to $u$.
\end{thm}
For a proof of this theorem see Lemma~10 in~\cite{Eder2013}. The formal statement of the theorem in Isabelle/HOL is as follows:
\begin{lstlisting}
lemma is-RB-upt-finite:
  assumes G $\subseteq$ %\RRm% and inj-on $\sig$ G and finite G
    and $\forall$g1$\in$G.%\,%$\forall$g2$\in$G. is-regular-spair g1 g2 $\longrightarrow$ $\sig$ (spair g1 g2) $\prect$ u $\longrightarrow$
			is-RB-in G ($\sig$ (spair g1 g2))
    and $\forall$i. i < length fs $\longrightarrow$ (1, i) $\prect$ u $\longrightarrow$ is-RB-in G (1, i)
  shows is-RB-upt G u
\end{lstlisting}
The second assumption of \textit{is-RB-upt-finite} merely expresses that the function \sig\ is
injective on $G$, that is, no two elements of $G$ have the same signatures.

Theorem~\ref{thm::Main} gives us some 
idea how to decide whether a given finite set $G$ is a rewrite basis up to $u$: 
it suffices to check the \emph{finitely many} signatures of regular S-pairs and canonical basis 
vectors. Note, however, that there is still an issue related to syzygy signatures: the definition 
of rewrite bases involves syzygy signatures, and deciding whether a given $u$ is a syzygy signature 
is a difficult problem---actually, as difficult as computing a Gr\"obner basis of the module of 
syzygies. Luckily, Theorem~\ref{thm::Main} does not only suggest a method for (semi-)deciding 
whether a given set is a rewrite basis, but it also gives rise to an algorithm for \emph{computing} 
rewrite bases which does not suffer from the problem with syzygy signatures just outlined. This 
algorithm is the subject of the next section.

% 3 pages

\section{Algorithms}
\label{sec::Algorithms}

As claimed above, Theorem~\ref{thm::Main} gives rise to an algorithm for computing rewrite bases, 
and in fact that algorithm bears close resemblance to Buchberger's algorithm for 
computing Gr\"obner bases: it is a critical-pair/completion algorithm that successively iterates 
through all S-pairs, applies a criterion for testing whether the S-pair under consideration must be 
reduced, \sig-reduces it to some normal form if necessary, and adds the result to the basis computed so far unless it be zero. Algorithm~\ref{alg::RB} summarizes the method just sketched in an 
imperative programming style; it is a slight variation of Algorithm~3 in~\cite{Eder2017}.
\begin{algorithm}
\caption{An algorithm for computing rewrite bases.}
\label{alg::RB}
\begin{algorithmic}[1]
\Require{sequence $(f_1,\ldots,f_m)$ of polynomials in $\RR$, admissible order $\preceq$ on $[X]$, 
compatible extension $\preceqt$ on $\TT$, rewrite order $\rw$}
\Ensure{rewrite basis $G$}
\Function{RB}{$(f_1,\ldots,f_m),\preceq,\preceqt,\rw$}
  \Let{$G$}{$\emptyset$}
  \Let{$S$}{$\{\sig(f_j\,\mathbf{e}_i-f_i\,\mathbf{e}_j)\ |\ 1\leq i<j\leq m\}$}
  \Let{$P$}{$\{\mathbf{e}_i\ |\ 1\leq i\leq m\}$}
  \While{$P\neq\emptyset$}\label{ln::while}
    \Let{$a$}{some element of $P$ with $\preceqt$-minimal signature}\label{ln::a}
    \Let{$P$}{$P\backslash\{a\}$}
    \If{$a=\mathbf{e}_i$ for some $i$}\label{ln::S1_0}
      \Let{$S$}{$S\cup\{\sig(f_i\,g-\overline{g}\,\mathbf{e}_i)\ |\ g\in G\}$}\label{ln::S1}
    \EndIf
    \If{$\neg$\textsc{sigCrit}($\rw,G,S,a$)}\label{ln::sigCrit}
      \Let{$b$}{result of regular \sig-reducing $a$ modulo $G$}
      \If{$\overline{b}=0$}\label{ln::zero?}
	\Let{$S$}{$S\cup\{\sig(b)\}$}\label{ln::S2}
      \Else
	\Let{$G$}{$G\cup\{b\}$}\label{ln::addG}
	\Let{$P$}{$P\cup\{\spair(g,b)\ |\ g\in G,\ \spair(g,b)\text{ is regular}\}$}
      \EndIf
    \EndIf
  \EndWhile
  \State \Return{$G$}\label{ln::return}
\EndFunction
\end{algorithmic}
\end{algorithm}

%\newpage
Several remarks on Algorithm~\ref{alg::RB} are in place:
\begin{itemize}
 \item The accumulator $G$ holds the basis computed so far, and $P$ is the set of elements 
that still have to be considered. It does not only contain regular S-pairs, but also the $m$ 
canonical basis vectors corresponding to the input-sequence $(f_1,\ldots,f_m)$. This justifies initializing $G$ by the empty set.

 \item The set $S$ contains the signatures of some known syzygies. It is initialized by the signatures of 
the \emph{Koszul syzygies} of the input sequence, and successively enlarged in Lines~\ref{ln::S1} 
and~\ref{ln::S2}. These syzygy-signatures are used to apply the syzygy criterion 
(Lemma~\ref{lem::SyzCrit}) in function \textsc{sigCrit}, see Algorithm~\ref{alg::sigCrit} below.

 \item It is important to note that in Line~\ref{ln::a} of Algorithm~\ref{alg::RB}, an element 
$a$ with minimal signature is taken from $P$. This is crucial for the correctness of the algorithm, 
since a different choice could lead to wrong results.

 \item Also note that $\sig(b)=\sig(a)$, since $b$ is the result of \emph{regular} \sig-reducing 
$a$, and regular \sig-reductions do not change signatures. This, together with the particular 
choice of $a$, implies that $G$ is computed by increasing signatures, i.\,e., the signatures of the 
elements $b$ added to $G$ in Line~\ref{ln::addG} are increasing.
\end{itemize}

Ignoring the \textsc{sigCrit}-test in Line~\ref{ln::sigCrit} of Algorithm~\ref{alg::RB} for the 
moment, the algorithm is partially correct. This follows from the fact that either 
$\overline{b}=0$, in which case $\sig(b)=\sig(a)$ is a syzygy signature, or $b$ is added to $G$, in 
which case it becomes the canonical rewriter in $\sig(b)=\sig(a)$ (this follows from the definition 
of rewrite orders) and is by construction regular top \sig-irreducible. Therefore, in either case 
the potentially enlarged set $G$ is a rewrite basis in $\sig(a)$ by 
Definition~\ref{dfn::RewriteBasis}, and upon termination of the algorithm, it is a rewrite basis in \emph{all} terms $u$ 
thanks to Theorem~\ref{thm::Main}.

The auxiliary function \textsc{sigCrit}, which is implemented in Algorithm~\ref{alg::sigCrit}, tests 
whether an S-pair $\spair(a,b)$ has to be \sig-reduced in Algorithm~\ref{alg::RB}. In a nutshell, 
it applies Lemma~\ref{lem::SyzCrit}, the syzygy criterion, and moreover checks whether the 
constituents of the S-pair are canonical rewriters in certain terms $u_a$ and $u_b$; if not, the 
S-pair does not have to be reduced, because either the canonical rewriters in these respective 
terms have been treated already, or will still be treated later on, and in either case there is 
nothing to be done for $\spair(a,b)$. There is one subtle point, though: Knowing that $\spair(a,b)$ 
is regular, one of $u_a$ or $u_b$ is strictly greater than the other by definition, and 
$\sig(\spair(a,b))=\max(u_a,u_b)$. W.\,l.\,o.\,g. assume $u_b\prect u_a$. So, by what has been 
said above, it should be clear that \textsc{sigCrit} is allowed to do the checks on 
$u_a=\sig(\spair(a,b))$ in Line~\ref{ln::u_a} of Algorithm~\ref{alg::sigCrit}, but it is perhaps 
not clear why the same checks may also be performed on the \emph{smaller} term $u_b$ that does not 
contribute to $\sig(\spair(a,b))$ at all. Indeed, answering this question is slightly intricate, 
and we confine ourselves here to pointing the interested reader to Lemma~12 in~\cite{Eder2013} for 
an explanation.
\begin{algorithm}
\caption{An algorithm for testing whether S-pairs must be regular \sig-reduced.}
\label{alg::sigCrit}
\begin{algorithmic}[1]
\Require{rewrite order $\rw$, $G\subseteq\RR^m$, $S\subseteq\TT$, regular $\spair(a,b)$ with 
$a,b\in G$}
\Ensure{`False' if $\spair(a,b)$ has to be regular \sig-reduced in Algorithm~\ref{alg::RB}}
\Function{sigCrit}{$\rw,G,S,\spair(a,b)$}
  \Let{$t$}{$\lcm(\lp(\overline{a}),\lp(\overline{b}))$}
  \Let{$u_a$}{$\frac{t}{\lp(\overline{a})}\sig(a)$}
  \Let{$u_b$}{$\frac{t}{\lp(\overline{b})}\sig(b)$}
  \If{$(\exists s\in S.\ s\,|\,u_a)\vee(\text{$a$ is not canonical rewriter in $u_a$ \wrt\ $G$})$}
      \label{ln::u_a}
    \State\Return{True}
  \EndIf
  \If{$(\exists s\in S.\ s\,|\,u_b)\vee(\text{$b$ is not canonical rewriter in $u_b$ \wrt\ $G$})$}
      \label{ln::u_b}
    \State\Return{True}
  \EndIf
  \State\Return{False}
\EndFunction
\end{algorithmic}
\end{algorithm}

We hope we could convince the reader about the partial correctness of Algorithms~\ref{alg::RB} 
and~\ref{alg::sigCrit} now; if not, a more thorough account on the whole subject can, as usual, be 
found in~\cite{Eder2017}. However, the algorithm is not only partially correct, but also 
terminates for every input; this claim will be investigated in Section~\ref{sec::Termination}. We 
summarize the result in a theorem:
\begin{thm}[Correctness of Algorithm~\ref{alg::RB}]
\label{thm::RB}
For every input, Algorithm~\ref{alg::RB} terminates and returns a rewrite basis $G$ \wrt\ 
$(f_1,\ldots,f_m)$, $\preceq$, $\preceqt$ and \rw. Furthermore, $\langle\overline{G}\rangle=\langle 
f_1,\ldots,f_m\rangle$.
\end{thm}

\begin{rem}
Algorithm~\ref{alg::sigCrit} corresponds to Algorithm~4 in~\cite{Eder2017}, which, however, is 
presented in a slightly different way. Namely, the two disjuncts in Lines~\ref{ln::u_a} 
and~\ref{ln::u_b} of Algorithm~\ref{alg::sigCrit} are combined into one single `rewritability' check 
in the cited article. This makes the formulation of the algorithm a bit more elegant.

Also, one has to take into account that the last argument of function \textsc{sigCrit} could be a 
canonical basis vector $\mathbf{e}_i$ rather than an S-pair. In that case, only the syzygy 
criterion is applied, i.\,e., $\exists s\in S.\ s\,|\,\mathbf{e}_i$.
\end{rem}

Let us now turn to the formalization of \textsc{RB} in Isabelle/HOL. There, it is natural to 
implement functions as \emph{functional} programs instead of imperative ones, so we define the 
tail-recursive function \textsf{rb-aux} for computing rewrite bases as follows:
\begin{lstlisting}
function rb-aux ::
      ((($\polymapping{\tau}$) list $\times$ $\tau$ list $\times$ ((($\polymapping{\tau}$) $\times$ ($\polymapping{\tau}$)) + nat) list) $\times$ nat) $\Rightarrow$
       ((($\polymapping{\tau}$) list $\times$ $\tau$ list $\times$ ((($\polymapping{\tau}$) $\times$ ($\polymapping{\tau}$)) + nat) list) $\times$ nat)
  where
    rb-aux ((gs, ss, []), z) = ((gs, ss, []), z) |
    rb-aux ((gs, ss, a # ps'), z) =
      (let ss' = new-syz-sigs ss gs a in
        if sig-crit gs ss' a %then%
          rb-aux ((gs, ss', ps'), z)
        else
          let b = sig-trd gs (poly-of-pair a) in
            if %\poly% b = 0 %then%
              rb-aux ((gs, ($\sig$ b) # ss', ps'), Suc z)
            else
              rb-aux ((b # gs, ss', add-spairs ps' gs b), z))
\end{lstlisting}
The function takes one argument, which in turn is a tuple consisting of four entries: a list $gs$ 
corresponding to the set $G$ in Algorithm~\ref{alg::RB}, a list $ss$ corresponding to $S$, a list 
$ps$ corresponding to $P$, and a natural number $z$ counting the total number of zero-reductions. 
The latter is a mere technicality only needed in Section~\ref{sec::ZeroReductions}, and may thus be ignored 
for the moment. The function not only returns $gs$, but also the other arguments, to facilitate 
formal reasoning about it---but of course only $gs$ is interesting from our perspective. Please 
note that the list $\fs$ and the various relations ($\preceq$, etc.) are still implicitly fixed in the 
theory context and therefore do not have to be passed as arguments to \textsf{rb-aux} explicitly.

The first part of the definition corresponds to the base case, where the list $ps$ is empty. 
The second part corresponds to the case where $ps$ contains at least one element, 
and can hence be decomposed into its head $a$ and tail $ps'$. Since we ensure that the list is 
always kept sorted by increasing signatures, $a$ is known to be an element with minimal 
signature, just as required in Line~\ref{ln::a} of Algorithm~\ref{alg::RB}. Then, $ss$ is enlarged 
by new syzygy-signatures in the auxiliary function \textsf{new-syz-sigs}, and the result is stored 
in $ss'$; this corresponds precisely to Lines~\ref{ln::S1_0} and~\ref{ln::S1} of 
Algorithm~\ref{alg::RB}. Afterward, the auxiliary function \textsf{sig-crit} is applied to $gs$, 
$ss'$ and $a$ to check whether $a$ has to be \sig-reduced or not. \textsf{sig-crit} is the 
formalization of function \textsc{sigCrit}, and since there is nothing special about its definition, 
we omit it here. Anyway, if \textsf{sig-crit} returns \textsf{True}, nothing needs to be done and 
\textsf{rb-aux} is called recursively on the remaining list $ps'$. Otherwise, $a$ is regular 
\sig-reduced to $b$ (taken care of by function \textsf{sig-trd}), and depending on whether $b$ is a 
syzygy or not its signature is added to $ss'$ or it is added to $gs$, and new S-pairs are added to 
$ps'$ by function \textsf{add-spairs}. So, in short, \textsf{rb-aux} corresponds exactly to 
Lines~\ref{ln::while}--\ref{ln::return} of Algorithm~\ref{alg::RB}. The remaining lines, 
corresponding to the initialization of $G$, $S$ and $P$, are covered by the way how the arguments of 
the initial call of \textsf{rb-aux} are constructed, as will be seen below.

Before, please note that the element-type of $ps$ is a \emph{sum type}, i.\,e., the disjoint 
union of two types: once the type of pairs of module elements, 
$(\polymapping{\tau})\times(\polymapping{\tau})$, and once the type \textsf{nat} of natural 
numbers. This is due to the fact that $ps$ may both contain S-pairs and canonical basis vectors: 
S-pairs are represented by the two elements they originate from, because these elements themselves 
are needed in \textsf{sig-crit}, and canonical basis vectors are compactly represented by their 
component, which is of course a natural number. Function \textsf{poly-of-pair} converts an object of 
this sum type into an actual module element of type $\polymapping{\tau}$, by either constructing 
an S-pair or returning a `full' basis vector.

The initial argument of \textsf{rb-aux} corresponds to the initial values of $G$, $S$ and $P$: $gs$ is the empty list, $ss$ is $\textsf{Koszul-syz-sigs}(\fs)$, which returns the signatures of the 
Koszul syzygies of \fs, and $ps$ is the list $\textsf{map}(\textsf{Inr},[0..<\textsf{length}(\fs)])$, 
representing the canonical basis vectors in the sum type mentioned above.

So, we can finally define function \textsf{rb} as follows:
\begin{lstlisting}
definition rb :: ($\polymapping{\tau}$) list $\times$ nat
  where rb = (let ((gs, _, _), z) =
		      rb-aux (([], Koszul-syz-sigs fs, map Inr [0..<length fs]), 0) 
		in (gs, z))
\end{lstlisting}
As can be seen, \textsf{rb} does not take any explicit arguments in the above definition, but it is 
implicitly parameterized over the constants fixed in the theory context (\fs, $\preceq$, $\preceqt$ 
and \rw).

In order to formally prove the correctness of \textsf{rb-aux}, and hence \textsf{rb}, we 
define an invariant \textsf{rb-aux-inv} of function \textsf{rb-aux} that holds for the initial 
argument, is preserved in every recursive call, and is strong enough to infer the desired 
properties of \textsf{rb} from it. Since the precise definition of the invariant is fairly lengthy, 
we only informally summarize its key characteristics here. $\textsf{rb-aux-inv}(gs,ss,ps)$ holds if
\begin{itemize}
 \item the signatures of the elements of $gs$ are strictly decreasing (note that new 
elements with larger signatures are added up front to $gs$),

 \item every element in $gs$ stems from regular \sig-reducing an S-pair of elements coming 
later in $gs$ (i.\,e., earlier during execution of the function), or from regular \sig-reducing a 
canonical basis vector,

 \item every element in $gs$ is regular \sig-irreducible modulo the elements coming after it in 
$gs$,

 \item every element of $gs$ belongs to the set \textsf{\RRm},

 \item $gs$ does not contain syzygies,

 \item for every $g$ in $gs$, the elements coming after it in $gs$ constitute a rewrite basis up to 
$\sig(g)$,
 
 \item every element in $ss$ is a syzygy signature,
 
 \item $ps$ is sorted by increasing signatures,
 
 \item no element in $ps$ has a signature which is strictly smaller than the signature of any 
element in $gs$, and

 \item $gs$ is a rewrite basis in all $\mathbf{e}_i$ which do not appear in $ps$ any more, and 
similar for S-pairs.
\end{itemize}
The first three items are only needed for proving termination of \textsf{rb-aux}, see 
Section~\ref{sec::Termination}. This list is not exhaustive; it is only meant to give an impression 
of how challenging it is to prove correctness of \textsf{rb-aux} and \textsf{rb} in a formal 
environment. In absolute figures, the whole proof, distributed across several lemmas, takes roughly 
1800 lines of Isabelle code---not counting the proofs of the necessary theoretical results shown in 
previous sections, like \textit{is-RB-upt-finite}. The claim that the invariant is 
preserved in the third recursive call of \textsf{rb-aux} turns out have the most difficult proof:
\begin{lstlisting}
lemma rb-aux-inv-preserved-3:
  fixes gs ss a ps
  defines ss' $\equiv$ new-syz-sigs ss gs a
  defines b $\equiv$ sig-trd gs (poly-of-pair a)
  assumes rb-aux-inv (gs, ss, a # ps) and $\neg$ sig-crit gs ss' a and %\poly% b $\neq$ 0
  shows rb-aux-inv (b # gs, ss', add-spairs ps gs b)
\end{lstlisting}

After having proved that \textsf{rb-aux-inv} holds for the initial argument of \textsf{rb-aux} and 
is preserved in each of the three recursive calls, and that \textsf{rb-aux} terminates (see 
Section~\ref{sec::Termination}), we can infer the following two key properties of \textsf{rb} which 
correspond to Theorem~\ref{thm::RB}:
\begin{lstlisting}
theorem rb-is-RB-upt: is-RB-upt (set (fst rb)) u%\\[-1ex]%
theorem ideal-rb-aux: ideal (%\poly% %\textasciigrave% set (fst rb)) = ideal (set fs)
\end{lstlisting}

\begin{rem}
Algorithm~\ref{alg::RB} and function \textsf{rb} could easily be adapted to not only compute a 
rewrite basis, and hence Gr\"obner basis of the ideal $\langle f_1,\ldots,f_m\rangle$, 
but also a Gr\"obner basis of the module of syzygies of $(f_1,\ldots,f_m)$. We do not consider this 
in the formalization, though.
\end{rem}

\subsection{Termination}
\label{sec::Termination}

Termination of the original $F_5$ algorithm had been an open problem for a long time, until it was 
eventually settled by~\cite{Galkin2012}. Later, \cite{Pan2012} proved termination of a more general 
signature-based algorithm, which happens to be equivalent to Algorithm~\ref{alg::RB}. The proof we 
modeled our formal Isabelle-proof after can be found in~\cite{Eder2013} (Theorem~20). Here, we 
present the key ideas of the proof, referring the interested reader to the cited article for more 
information about it.

Assume $(g_1,g_2,g_3,\ldots)$ is the sequence of elements added to $G$ by Algorithm~\ref{alg::RB}, 
in that order. We want to show that this sequence is finite. First, introduce the following 
relation $\sim$ on $\RR^m$: $a\sim b\ :\Leftrightarrow\ 
\lp(\overline{b})\sig(a)=\lp(\overline{a})\sig(b)$. $\sim$ is an equivalence relation, and 
therefore allows one to partition the sequence into subsets of equivalent elements \wrt\ $\sim$. 
Next, one can prove that only finitely many of these subset are non-empty, using 
Noetherianity of $\RR^m$ and further properties of the sequence that follow from its being 
constructed by Algorithm~\ref{alg::RB}, e.\,g., no element is regular \sig-reducible by the others. 
Finally, one can prove by induction on the finitely many non-empty sets $R$ that each of them
is finite, because every element of $R$ corresponds to an S-pair of elements in `previous' sets, which are finite by the induction hypothesis. This concludes the proof.

\begin{rem}
Readers not so familiar with signature-based algorithms might wonder why the well-known termination 
proof of Buchberger's algorithm does not work for signature-based algorithms. The reason is simple: 
a new element $b$ added to the basis is only regular \sig-irreducible, which unfortunately does not 
imply that $\overline{b}$ is irreducible in the traditional sense of polynomial reduction. In 
particular, $\lp(\overline{b})$ might even be divisible by $\lp(\overline{g})$ for some $g$ in the 
current basis---something which cannot happen in Buchberger's algorithm, which in turn is what 
the termination proof of Buchberger's algorithm mainly rests upon.
\end{rem}

In the formalization, the theorem needed for establishing termination of function \textsf{rb-aux} 
is as follows:
\begin{lstlisting}
lemma rb-termination:
  fixes seq :: nat $\Rightarrow$ ($\polymapping{\tau}$)
  assumes $\forall$i j. i < j $\longrightarrow$ $\sig$ (seq i) $\prect$ $\sig$ (seq j)
    and $\forall$i. ($\exists$j<length fs. $\sig$ (seq i) = (0, j) $\wedge$ lp (%\poly% (seq i)) $\preceq$ lp (fs ! j)) $\vee$
             ($\exists$j k. is-regular-spair (seq j) (seq k) $\wedge$
		    %\poly% (spair (seq j) (seq k)) $\neq$ 0 $\wedge$
		    $\sig$ (seq i) = $\sig$ (spair (seq j) (seq k)) $\wedge$
		    lp (%\poly% (seq i)) $\preceq$ lp (%\poly% (spair (seq j) (seq k))))
    and $\forall$i. $\neg$ is-sig-red ($\prect$) ($\preceq$) (seq %\textasciigrave% {0..<i}) (seq i)
    and range seq $\subseteq$ %\RRm% and 0 $\notin$ %\poly% %\textasciigrave% range seq
    and $\forall$i. is-sig-GB-upt (seq %\textasciigrave% {0..<i}) ($\sig$ (seq i))
  shows False
\end{lstlisting}
So, we assume that there exists an infinite sequence $seq$ with the listed properties and derive a 
contradiction; hence, any such sequence must be finite. $seq$ is modeled as a function from the 
natural numbers to module elements of type $\polymapping{\tau}$, which means that the $i$-th 
element of $seq$ is simply $seq(i)$ and the set of all elements of $seq$ is $\textsf{range}(seq)$. 
A close inspection of the presumed properties of $seq$ reveals that they essentially correspond to 
the first six properties of $gs$ in
the above list characterizing \textsf{rb-aux-inv}. The only real difference is that the order of the
elements in $seq$ corresponds to the order in which they are generated by function \textsf{rb-aux},
which is the \emph{reversed} order compared to $gs$. This explains why, for instance, the signatures
in $seq$ must be strictly increasing, whereas in $gs$ they must be strictly decreasing. 

From \textit{rb-termination} we can conclude that function \textsf{rb-aux} terminates for all 
arguments satisfying the invariant \textsf{rb-aux-inv}, which in particular includes the initial 
argument specified by function \textsf{rb}. This finishes the proof of total correctness of 
\textsf{rb}.

% 5.5 pages

\section{Optimality Results}
\label{sec::Optimality}

\subsection{No Zero-Reductions}
\label{sec::ZeroReductions}

The original goal of signature-based algorithms is to detect and avoid as many useless 
zero-reductions as possible, and thus speed up the computation of Gr\"obner bases. Practical 
experience shows that this goal is indeed achieved (see Section~\ref{sec::Computations}), and 
theory even tells us that in some situations zero-reductions can be avoided altogether:
\begin{thm}
\label{thm::ZeroReductions}
Let $(f_1,\ldots,f_m)$ be a regular sequence and assume $\preceqt\ =\ \preceqpot$, i.\,e., 
$\preceqt$ is a POT-extension of $\preceq$. Then Algorithm~\ref{alg::RB} does not \sig-reduce any 
element to zero, meaning that the test in Line~\ref{ln::zero?} of that algorithm always yields 
`False'.
\end{thm}
The proof of this celebrated result, which is presented as Corollary~7.1 
in~\cite{Eder2017}, is actually not very difficult. It proceeds along the following 
lines: Using $\preceqpot$, the rewrite basis is computed incrementally, i.\,e., first for $(f_1)$, 
then for $(f_1,f_2)$, and so on. The sequence $(f_1,\ldots,f_m)$ being regular implies that the 
only syzygies $a$ satisfying $\sig(a)=t\,\mathbf{e}_i$, for $1\leq i\leq m$ and $t\in [X]$, are in 
the module of principal syzygies of $(f_1,\ldots,f_i)$---a generating set of which is added to $S$ 
in Line~\ref{ln::S1}. However, every zero-reduction corresponds to precisely such a syzygy, and 
therefore is detected beforehand by the syzygy criterion implemented in function \textsc{sigCrit}. It should be observed that \cite{Eder2017} need the additional assumption that \rw\ be either 
\rwrat\ or \rwadd. We do not need this assumption because of our slightly different implementation 
of function \textsc{sigCrit}.

The formalization of Theorem~\ref{thm::ZeroReductions} in Isabelle/HOL begins with the definition 
of regular sequences:
\begin{lstlisting}
definition is-regular-sequence :: ($\polymapping{\alpha}$) list $\Rightarrow$ bool
  where is-regular-sequence fs $\longleftrightarrow$
		      ($\forall$j<length fs. $\forall$q. q * fs ! j $\in$ ideal (set (take j fs)) $\longrightarrow$
					  q $\in$ ideal (set (take j fs)))
\end{lstlisting}
As can be seen, \textsf{is-regular-sequence} is a predicate on lists of polynomials. The definition 
avoids any reference to quotient rings by unfolding the definition of zero-divisors in such rings. 
Function $\textsf{take}(j,\fs)$ returns the list of the first $j$ elements of \fs.

Proving that there are no zero-reductions in function \textsf{rb} obviously boils down to 
proving that the second case in the second part in the definition of \textsf{rb-aux} cannot occur. 
This means that whenever \textsf{sig-crit} fails to hold for some $a$, the result of regularly 
\sig-reducing $a$ cannot be a syzygy:
\begin{lstlisting}
lemma rb-aux-inv2-no-zero-red:
  assumes is-regular-sequence fs and is-pot-ord
    and rb-aux-inv2 (gs, ss, a # ps) and $\neg$ sig-crit gs (new-syz-sigs ss gs a) a
  shows %\poly% (sig-trd gs (poly-of-pair a)) $\neq$ 0
\end{lstlisting}
Here, \textsf{is-pot-ord} expresses the fact that the implicitly fixed order $\preceqt$ is a 
POT-extension of $\preceq$. The predicate \textsf{rb-aux-inv2} is a strengthened version of \textsf{rb-aux-inv}, 
which can also be proved to be an invariant of \textsf{rb-aux} if \fs\ is a regular sequence and 
\textsf{is-pot-ord} holds. It additionally requires $ss$ to contain all necessary 
syzygy-signatures, something which is not needed for proving correctness of \textsf{rb-aux} and 
hence is not encoded in \textsf{rb-aux-inv}.

As a consequence of \textit{rb-aux-inv2-no-zero-red} and the fact that \textsf{rb-aux-inv2} holds 
for the initial argument of \textsf{rb-aux} as specified in \textsf{rb}, we can infer that indeed no 
zero-reductions take place. This result is formulated using the second return value, $z$, of 
\textsf{rb}, which counts the total number of zero-reductions:
\begin{lstlisting}
corollary rb-aux-no-zero-red:
  assumes is-regular-sequence fs and is-pot-ord
  shows snd rb = 0
\end{lstlisting}

\subsection{Minimal Signature Gr\"obner Bases}
\label{sec::Minimality}

Just as traditional Gr\"obner bases, signature Gr\"obner bases are not unique. Hence, we can 
define \emph{minimal} signature Gr\"obner bases as follows:
\begin{dfn}[Minimal Signature Gr\"obner Basis]
\label{dfn::MinSigGB}
A signature Gr\"obner basis is called \emph{minimal} if, and only if, none of its elements is top 
\sig-reducible modulo the other elements.
\end{dfn}
Note that minimal signature Gr\"obner bases have nothing to do with minimal Gr\"obner bases in the usual sense: if $G$ 
is a minimal signature Gr\"obner basis, then $\overline{G}$ is not automatically a minimal 
Gr\"obner basis, that is, there could exist $p_1,p_2\in\overline{G}$ with $p_1\neq p_2$ and 
$\lp(p_1)\,|\,\lp(p_2)$. Nevertheless, minimal signature Gr\"obner bases deserve the name, since every ideal $I\subseteq\RR^m$ has one \emph{unique} minimal signature Gr\"obner basis $G$, and moreover any other signature Gr\"obner basis $H$ of $I$ satisfies $\{\sig(g)\,|\,g\in G\}\subseteq\{\sig(h)\,|\,h\in H\}$ and $\{\lp(\overline{g})\,|\,g\in G\}\subseteq\{\lp(\overline{h})\,|\,h\in H\}$; see Lemma~4.3 
in~\cite{Eder2017} for details.

Surprisingly, when using \rwrat\ as the rewrite order, \textsf{rb-aux} automatically computes 
minimal signature Gr\"obner bases (recall from Proposition~\ref{prop::RB} that rewrite bases are 
also signature Gr\"obner bases). The following theorem corresponds to Corollary~7.3 
in~\cite{Eder2017}:
\begin{thm}
\label{thm::MinSigGB}
Assume $\rw\ =\ \rwrat$. Then the rewrite basis computed by Algorithm~\ref{alg::RB} is also a 
minimal signature Gr\"obner basis.
\end{thm}
Therefore, \rwrat\ is the optimal rewrite order in terms of the size of the resulting basis and 
the number of S-pairs that must be dealt with. Still, as noted in point (c) of Section~14.3 
in~\cite{Eder2017}, other rewrite orders, such as \rwadd, can lead to a comparable overall performance
of the algorithm.

Again, Definition~\ref{dfn::MinSigGB} and
Theorem~\ref{thm::MinSigGB} translate naturally into in Isabelle/HOL, as shown below:
\begin{lstlisting}
definition is-min-sig-GB :: ($\polymapping{\tau}$) set $\Rightarrow$ bool
  where is-min-sig-GB G $\longleftrightarrow$
		    G $\subseteq$ %\RRm% $\wedge$ ($\forall$u. snd u < length fs $\longrightarrow$ is-sig-GB-in G u) $\wedge$
		    ($\forall$g$\in$G. $\neg$ is-sig-red ($\preceqt$) (=) (G - {g}) g)%\\[-1ex]%
corollary rb-aux-is-min-sig-GB:
  assumes ($\rw$) = ($\rwrat$)
  shows is-min-sig-GB (set (fst rb))
\end{lstlisting}

% 2.5 pages

\section{Code Generation and Computations}
\label{sec::Computations}

When it comes to actually computing rewrite bases, the following two observations are important:
\begin{itemize}
 \item Algorithm~\ref{alg::RB} and function \textsf{rb} operate on module elements in $\RR^m$, or 
objects of type $\polymapping{\tau}$, respectively. Operations on such objects, such as addition, 
multiplication, etc., are of course $m$-times more expensive than on ordinary polynomials in $\RR$.

 \item A close investigation of said algorithms and their sub-algorithms, such as \emph{regular} 
\sig-reduction, reveals that in fact only the signature $\sig(a)$ and the polynomial part 
$\overline{a}$ of module elements $a\in\RR^m$ must be known for executing the algorithms. 
Therefore, the whole computation of rewrite bases can be made more efficient by letting the 
functions operate on sig-poly-pairs (see Definition~\ref{dfn::SPP}) instead of full module elements.
\end{itemize}
In the formalization, we take the preceding observations into account by \emph{refining} function 
\textsf{rb} and all other functions it depends on to new functions that operate on sig-poly-pairs, 
i.\,e., objects of type $\tau\times(\polymapping{\alpha})$. Of course, we formally prove that the 
refined functions behave precisely as the original ones and therefore inherit all their main 
properties. Eventually we end up with a function \textsf{gb-sig} that takes a list of polynomials 
as input, employs the refined version of \textsf{rb-aux} (called \textsf{rb-spp-aux}) for computing 
a rewrite basis of it, which is a list of sig-poly-pairs. Finally, it projects the elements of 
this list onto their second entries to obtain again a list of polynomials which constitute a 
Gr\"obner basis of the input. Furthermore, \textsf{gb-sig} is parameterized over $\preceq$, 
$\preceqt$ and \rw.

Thanks to Isabelle's \emph{code generator}, the mechanically verified function \textsf{gb-sig} can be 
used to effectively compute Gr\"obner bases. In a nutshell, this works by translating the 
definitions of \textsf{gb-sig} and its sub-algorithms, which are universally quantified equalities 
in Isabelle/HOL, into operationally equivalent procedures operating on concrete data structures in 
SML, OCaml, Scala or Haskell. The translation is implemented in such a way that the generated 
executable programs can be trusted to inherit all correctness properties of the abstract 
Isabelle-functions. More information about code generation in Isabelle can be found 
in~\cite{Haftmann2013,Haftmann2018}.

In our concrete case, multivariate polynomials are represented efficiently as ordered (\wrt\ 
$\preceq$) associative lists, mapping power-products to coefficients. This formally verified 
concrete representation, which is part of~\cite{Sternagel2010}, allows us to provide efficient 
implementations of all frequently used operations, e.\,g., addition, \textsf{lp}, etc.

A typical invocation of \textsf{gb-sig} within Isabelle, which automatically triggers code 
generation into SML and execution of the resulting program, could look as follows:\footnote{The 
suffixes `-pprod' are technical artifacts that can safely be ignored here.}
\begin{lstlisting}
value [code] gb-sig-pprod (POT DRLEX) rw-rat-strict-pprod
		[X ^ 2 * Z ^ 3 + 3 * X ^ 2 * Y, X * Y * Z + 2 * Y ^ 2]
\end{lstlisting}
This instruction immediately returns the following $4$-element Gr\"obner basis, computed 
over the field of rational numbers \wrt\ the POT extension of the degree-reverse-lexicographic 
ordering and rewrite order \rwrat:
\begin{lstlisting}
[(3 / 4) * X ^ 3 * Y ^ 2 - 2 * Y ^ 4, - 4 * Y ^ 3 * Z - 3 * X ^ 2 * Y ^ 2,
    X * Y * Z + 2 * Y ^ 2, X ^ 2 * Z ^ 3 + 3 * X ^ 2 * Y]
\end{lstlisting}
The auxiliary constants \textsf{X}, \textsf{Y} and \textsf{Z} are introduced for 
conveniently writing down trivariate polynomials; further indeterminates can easily 
be added on-the-fly, without even having to adapt the underlying type. More sample computations can 
be found in theory \textit{Signature-Examples} of the formalization.

Besides simple examples as the one shown above, \textsf{gb-sig} can also be tested on common 
benchmark problems and compared to other implementations of Gr\"obner bases. 
Table~\ref{tab::Timings} shows such a comparison to a formally verified implementation of 
Buchberger's algorithm with product- and chain-criterion in Isabelle/HOL, called 
\textsf{gb} and described in~\cite{Maletzky2018}, and to function \textsf{GroebnerBasis} in 
\textit{Mathematica}~11.3. Since this article is not meant as an exhaustive survey on the efficiency 
of different Gr\"obner basis algorithms, we confine ourselves here to present results of 
computations over the rationals \wrt\ the POT extension of the degree-reverse-lexicographic ordering 
and rewrite order \rwrat. We shall emphasize, however, that other order relations and rewrite 
orders are formalized, too, and may hence be used in computations without further ado.

\begin{table}[t]
\centering
\caption{Timings (in seconds) and total number of zero-reductions of Gr\"obner basis 
computations.\newline `?' indicates that the computation was aborted after $1200$ seconds.}
\label{tab::Timings}
\begin{tabular}{l r r r r r}
 & \multicolumn{2}{c}{\textsf{gb-sig}} & 
\multicolumn{2}{c}{\textsf{gb}} & \hspace{1em}\textit{Mathematica}\\\cline{2-6}
Benchmark\hspace{2em} & Time & \#0-red & \hspace{2em}Time & \#0-red & Time\\\hline
cyclic-5 & 0.1 & 0 & 0.1 & 79 & 0.0\\
cyclic-6 & 2.0 & 8 & 186.2 & 517 & 0.3\\
cyclic-7 & 544.7 & 36 & ? & ? & ?\\
katsura-6 & 0.9 & 0 & 9.5 & 159 & 0.5\\
katsura-7 & 22.4 & 0 & 270.0 & 355 & 3.7\\
katsura-8 & 1005.4 & 0 & ? & ? & 42.0\\
eco-9 & 3.0 & 0 & 24.2 & 685 & 2.8\\
eco-10 & 32.0 & 0 & 255.7 & 1572 & 27.9\\
eco-11 & 297.2 & 0 & ? & ? & 263.0\\
noon-5 & 0.4 & 0 & 0.5 & 208 & 0.1\\
noon-6 & 8.7 & 0 & 13.8 & 738 & 1.0\\
noon-7 & 213.5 & 0 & 289.2 & 2467 & 12.4
\end{tabular}
\end{table}

\begin{rem}
The timings for \textit{Mathematica} have to be read with care: \textit{Mathematica} always 
computes a reduced Gr\"obner basis, whereas the results returned by \textsf{gb-sig} and 
\textsf{gb} are not necessarily reduced. So, the timings of \textit{Mathematica} must be understood 
as a mere reference mark for highly sophisticated, state-of-the-art computer algebra software. 
It is not surprising that our formally verified function \textsf{gb-sig} cannot compete with it in 
most cases.
\end{rem}

% 2 pages

\section{Conclusion}
\label{sec::Conclusion}
% 1.5 pages

In this paper we presented a formalization of signature-based algorithms for computing 
Gr\"obner bases in Isabelle/HOL. The formalization is generic, executable, and covers not only 
correctness but also optimality (no zero-reductions, minimal signature Gr\"obner bases) of the 
implemented algorithms.

The formalization effort was roughly three months of full-time work. This might not sound very 
much, but it must once again be noted that we could make heavy use of existing formalizations of 
multivariate polynomials and modules thereof, as well as Gr\"obner bases theory, in Isabelle/HOL. 
Otherwise, it would have taken a lot longer. The total number of lines of code is $\sim 11440$, 
distributed over the five theories \textit{Prelims} (general facts about lists, relations, etc.; 
$\sim 960$ lines), \textit{More-MPoly} (general properties of polynomials; $\sim 440$ lines), 
\textit{Quasi-PM-Power-Products} (facts about power-products; $\sim 290$ lines), 
\textit{Signature-Groebner} (main theory; $\sim 9370$ lines) and \textit{Signature-Examples} (code 
generation and sample computations; $\sim 380$ lines). Proofs are intentionally given in a quite 
verbose style for better readability.

\subsection{Related Work}
\label{sec::Related}

Even though signature-based algorithms have, to the best of our knowledge, not been formalized in 
any other proof assistant so far, formalizations of traditional Gr\"obner bases theory exist in 
various systems.

The first formalization of Gr\"obner bases dates back to~\cite{Thery2001} and~\cite{Persson2001} in 
the Coq proof assistant (\cite{Bertot2004}). Later,~\cite{Schwarzweller2006} formalized the purely 
theoretical aspects of the theory in Mizar (\cite{Bancerek2015}). \cite{Jorge2009} 
and~\cite{Medina-Bulo2010} implemented formally verified versions of 
Buchberger's algorithm in OCaml and Common LISP, respectively; the former was verified using Coq, 
and the latter using ACL2 (\cite{Kaufmann2000}). And, of course, the work presented in this paper 
heavily rests on the formalization of traditional Gr\"obner bases theory by~\cite{Immler2016} in 
Isabelle/HOL.

\cite{Buchberger2004} and~\cite{Craciun2008} took a slightly different approach: they managed to 
automatically synthesize Buchberger's algorithm from a formal description of its specification in 
the Theorema system (\cite{Buchberger2016}). In the same system, we formalized a generalization of 
Gr\"obner bases to \emph{reduction rings} (\cite{Maletzky2016}).

Finally, it must also be mentioned that Gr\"obner bases methodology for a long time has been, and 
still is, successfully applied in automated theorem proving, as a black-box algorithm for proving 
universal equalities and inequations over algebraically closed fields; see for 
instance~\cite{Harrison2001} and~\cite{Chaieb2007}.

\subsection{Future Work}
\label{sec::FutureWork}

The present formalization could be extended in several ways. First of all, function \textsf{gb-sig} 
could be improved by \emph{inter-reducing} intermediate bases when $\preceqpot$ is used as the 
module term order. This idea, due to~\cite{Stegers2006,Eder2010}, has the potential of speeding up 
computations, but inter-reducing intermediate bases turns out to be much more subtle in the 
signature-based setting than it is in the traditional setting.

Another possible improvement of \textsf{gb-sig} consists of implementing the $F_4$-style 
reduction, as proposed by~\cite{Faugere1999}. This approach not only \sig-reduces one polynomial at 
a time, but several polynomials simultaneously by row-reducing certain matrices. Incidentally, the 
$F_4$ algorithm and corresponding $F_4$-style reduction are part of the formalization 
by~\cite{Immler2016} (described in~\cite{Maletzky2018}), and therefore could be incorporated into 
the formalization presented here with only moderate effort. The main reason why we have not done so 
as of yet is that no increase in performance can be expected from it in this concrete case: 
matrices are represented densely as immutable arrays in Isabelle/HOL, but $F_4$-style reductions 
only make sense if sparse matrices are stored \emph{efficiently}, possibly 
even involving some sort of compression. Formalizing better representations of sparse matrices in 
Isabelle/HOL is left for future work.

A third potential improvement of the efficiency of the algorithms is the use of more sophisticated 
data-structures, e.\,g.\ tournament trees, kd-trees, and others. \cite{Roune2012} review some of these 
data-structures and how they can reasonably be used in the computation of Gr\"obner bases by 
signature-based algorithms.

\paragraph*{Acknowledgments}
I thank the anonymous reviewers for their valuable remarks.

%\section*{References}
\bibliographystyle{elsarticle-harv}
\bibliography{Paper}

\newpage
\appendix
\section{Translation between Mathematics and Formalization}
\label{apx::Translation}

Table~\ref{tab::Translation} lists several concepts of the theory and how they translate 
into our formalization as presented in this exposition, and into the actual Isabelle sources of the 
formalization. Differences between the latter two stem from increasing the readability of the paper 
and have no deeper significance; in fact, readers not intending to look at the Isabelle sources may 
safely ignore the last column.

\begin{table}[t]
\centering
\caption{Translations of concepts between informal mathematics, the formalization as presented in 
this paper, and the actual Isabelle sources of the formalization.}
\label{tab::Translation}
\begin{tabular}{l l l}
Mathematics\hspace{4em} & Formalization (paper) & Formalization (sources)\\\hline
$\KK$ & $\beta$ & $\beta$\\
$\RR$ & \polymapping{\alpha} & \polymapping{\alpha}\\
$\RR^m$ & \polymapping{\tau};\quad\textsf{\RRm} & \polymapping{\tau};\quad \textsf{dgrad-sig-set}\\
$\langle\cdot\rangle$ & \textsf{ideal} & \textsf{ideal}\\
$(f_1,\ldots,f_m)$ & \textsf{fs} & \textsf{fs}\\
$\overline{a}$ & \textsf{\poly} $a$ & \textsf{rep-list} $a$\\
\supp & \textsf{supp} & \textsf{keys}\\
\coeff & \textsf{coeff} & \textsf{lookup}\\
$\preceq$ & $\preceq$ & $\preceq$\\
$\preceqt$ & $\preceqt$ & $\preceqt$\\
\lp & \textsf{lp} & \textsf{punit.lt}\\
\lc & \textsf{lc} & \textsf{punit.lc}, \textsf{lc}\\
\sig & \sig & \textsf{lt}\\

$\red{\mathrm{r}_1}{\mathrm{r}_2}{G}$ & \textsf{sig-red} $\mathrm{r}_1$ $\mathrm{r}_2$ $G$ & 
\textsf{sig-red} $\mathrm{r}_1$ $\mathrm{r}_2$ $G$\\
$t\,u$\quad($t\in [X]$, $u\in\TT$) & $t \otimes u$ & $t\oplus u$\\
$u\,|\,v$\quad($u,v\in\TT$) & $u$ \dvdt\ $v$ & $u\ \mathsf{adds}_\mathsf{t}\ v$\\
$c\,t\,a$\quad($c\in\KK$, $t\in [X]$, $a\in\RR^m$) & \textsf{monom-mult} $c$ $t$ $a$ & 
\textsf{monom-mult} $c$ $t$ $a$\\
$(\sig(a),\overline{a})$ & \textsf{spp-of} $a$ & \textsf{spp-of} $a$\\
$\rwrat$ & $\rwrat$ & \textsf{rw-rat}\\
$\rwadd$ & $\rwadd$ & \textsf{rw-add}
\end{tabular}
\end{table}

% \newpage
% \section*{Curriculum Vitae}
% 
% I gained a Bachelor's degree in technical mathematics from Johannes Kepler University (JKU) Linz, 
% Austria, in 2011. Afterward, I also did a Master's in computer mathematics 
% at the same university, finishing in 2013. From 2013 to 2016 I was a PhD student under the 
% supervision of Bruno Buchberger at the Research Institute for Symbolic Computation (RISC), JKU 
% Linz, where my work focused on the formalization of reduction ring theory in the Theorema proof 
% assistant. Since fall 2016 I am a post-doctoral researcher at RISC, working on the formalization of 
% topics related to Gr\"obner bases in Isabelle/HOL and Theorema.
\end{document}